\documentclass[11pt]{article}
\usepackage[T2A,T1]{fontenc}
\usepackage[cp1250]{inputenc}
\usepackage{amsmath} 
\usepackage{amssymb} 
\usepackage{a4wide} 
\usepackage[left=3cm]{geometry} 
\usepackage{braket}
\usepackage{slashed}
\usepackage{bbm}
\usepackage{comment}
\usepackage{yfonts}
\usepackage{simplewick}
\usepackage[boxsize=0.7em,centertableaux]{ytableau}
\usepackage{empheq}
\usepackage{tikz}
\usetikzlibrary{matrix,arrows}
\usetikzlibrary{decorations.markings, decorations.pathreplacing}

\allowdisplaybreaks

\numberwithin{equation}{section}

\title{On even spin $\mathcal{W}_{\infty}$}
\author{Tomáš Procházka}

\begin{document}

\bibliographystyle{hieeetr}

\vskip 2.1cm

\centerline{\large \bf On even spin $\mathcal{W}_{\infty}$}
\vspace*{8.0ex}

\centerline{\large \rm Tom\'{a}\v{s} Proch\'{a}zka\footnote{Email: {\tt tomas.prochazka@lmu.de}}}

\vspace*{8.0ex}

\centerline{\large \it Arnold Sommerfeld Center for Theoretical Physics}
\centerline{\large \it Ludwig Maximilian University of Munich}
\centerline{\large \it Theresienstr. 37, D-80333 München, Germany}
\vspace*{2.0ex}

\vspace*{6.0ex}

\centerline{\bf Abstract}
\bigskip

We study the even spin $\mathcal{W}_\infty$ which is a universal $\mathcal{W}$-algebra for orthosymplectic series of $\mathcal{W}$-algebras. We use the results of Fateev and Lukyanov to embed the algebra into $\mathcal{W}_{1+\infty}$. Choosing the generators to be quadratic in those of $\mathcal{W}_{1+\infty}$, we find that the algebra has quadratic operator product expansions. Truncations of the universal algebra include principal Drinfe\v{l}d-Sokolov reductions of $BCD$ series of simple Lie algebras, orthogonal and symplectic cosets as well as orthosymplectic $Y$-algebras of Gaiotto and Rap\v{c}\'{a}k. Based on explicit calculations we conjecture a complete list of co-dimension $1$ truncations of the algebra.

 \vfill \eject

\tableofcontents

\setcounter{footnote}{0}

\newpage

\section{Introduction}
$\mathcal{W}$-algebras and their incarnations as affine Yangians \cite{Maulik:2012wi,tsymbaliuk2017affine}, degenerate double affine Hecke algebras \cite{schifvas} or cohomological Hall algebras \cite{Kontsevich:2010px} plan an important role in various areas of mathematical physics. They were originally introduced in the context of integrable hierarchies of partial differential equations and soon after in conformal field theory. Some more recent applications include four-dimensional $\mathcal{N}=2$ gauge theories \cite{Alday:2009aq,Beem:2013sza,Gaiotto:2017euk}, $M$-theory \cite{Costello:2017fbo,Gaiotto:2019wcc} or higher spin $AdS_3/CFT_2$ dualities \cite{Gaberdiel:2010pz}.

One of the most exciting $\mathcal{W}$-algebras is the universal two-parametric family of algebras called $\mathcal{W}_{1+\infty}$ which has one generating field of every integer spin. It interpolates between algebras of $\mathcal{W}_N$ family which are among the most well studied examples of $\mathcal{W}$-algebras. The very distinct property of $\mathcal{W}_{1+\infty}$ is that it has quantum triality symmetry \cite{Gaberdiel:2011wb}. The vacuum character of this algebra is given by MacMahon function which connects the representation theory of $\mathcal{W}_{1+\infty}$ to combinatorics of plane partition, plane tilings and dimer models.

A less studied example of universal interpolating algebra is the even spin $\mathcal{W}_\infty$ which is freely generated by fields of every even spin \cite{Gaberdiel:2011nt,Ahn:2011pv,Candu:2012ne,Kanade:2018qut}. Here we want to study this algebra in more detail. We first review and slightly extend the primary bootstrap approach to this algebra \cite{Candu:2012ne}. A different choice of independent structure constants with respect to \cite{Candu:2012ne} eliminates ambiguities arising from square root factors and spurious duality symmetries. Using three elementary minimal representations of the algebra we introduce a convenient triality covariant parametrization analogous to parametrization of $\mathcal{W}_\infty$ introduced in \cite{Prochazka:2014gqa}. We also introduce another parametrization which is closer to parametrization of Gaiotto and Rap\v{c}\'{a}k \cite{Gaiotto:2017euk} featuring the Kapustin-Witten parameter. Next we identify our parameters with parameters of well-known truncations of the algebra, including orthogonal and symplectic quotients and principal Drinfeld-Sokolov reductions of $BCD$ families of simple Lie algebras. We also compare even spin $\mathcal{W}_\infty$ with orthosymplectic $Y$-algebras introduced in \cite{Gaiotto:2017euk}.

In the next section we use the results of Fateev and Lukyanov \cite{Lukyanov:1989gg} to embed even spin $\mathcal{W}_{\infty}$ into $\mathcal{W}_{1+\infty}$. Although in retrospect this is not very surprising, the possibility of doing this at the level of non-linear quantum algebras it is not at all obvious. Actually the principal Drinfe\v{l}d-Sokolov reductions of all simple Lie algebras are subalgebras of truncations of $\mathcal{W}_{1+\infty}$ (except for $F_4$ where this is not known) so in this sense $\mathcal{W}_{1+\infty}$ can be thought of as interpolating algebra for all $\mathcal{W}$-algebras associated to simple Lie algebras via Drinfe\v{l}d-Sokolov reduction (except for $F_4$). One would like to start with Miura operator for $GL(N)$ $\mathcal{W}$-algebras and fold it to obtain a Miura operator for $BCD$-type algebras \cite{drinfeld1985lie}. Unfortunately it is not clear how to do this at quantum level. The trick used by Fateev and Lukyanov is instead to consider $D_n$ algebras which have additional Pfaffian generator of dimension $n$ and study its operator product expansion with itself. From here we can identify the generators of even spin $\mathcal{W}_\infty$ as quadratic composites of the generators of $\mathcal{W}_{1+\infty}$. We verify by explicit calculations using $\mathtt{OPEdefs}$ \cite{Thielemans:1991uw} that the resulting subalgebra quadratically closes (up to sum of spins $20$) and we find a map between parameters of $\mathcal{W}_{1+\infty}$ and those of even spin $\mathcal{W}_\infty$. The operator product expansions in the quadratic basis share many nice properties with $\mathcal{W}_{1+\infty}$ or even the matrix valued $\mathcal{W}_{1+\infty}$ \cite{Eberhardt:2019xmf}, but unlike in those cases it is not clear at the moment how to sum the derivative terms.

In the following section we list the known truncation curves and study co-dimension $1$ truncations of vacuum representation up to level $12$. The structure seems to be more complicated than in the case of $\mathcal{W}_\infty$, but all truncations found agree nicely with a simple formula for truncation curves and also with Gaiotto-Rap\v{c}\'{a}k $Y$-algebras. We conjecture a general formula for the level of the first singular vector in all of these truncations. In the last section we show on an example of $\mathfrak{so}(2n+1)_k$ that the gluing procedure of \cite{Prochazka:2017qum} generalizes also to the orthosymplectic case.

\section{Primary bootstrap}
Let us review and slightly extend the results of \cite{Candu:2012ne} where the authors used the OPE bootstrap to construct even spin $\mathcal{W}_\infty$ in the primary basis\footnote{The bootstrap procedure is mathematically formalized in \cite{de2005freely} and recently reviewed in \cite{Linshaw:2017tvv}.}. We assume to have one generating field of each even spin with the following operator product expansions between the fields up to sum of spins $14$
\begin{align}
\nonumber
W_4 W_4 & \sim C_{44}^0 \mathbbm{1} + C_{44}^4 W_4 + C_{44}^6 W_6 \\
\nonumber
W_4 W_6 & \sim C_{46}^4 W_4 + C_{46}^6 W_6 + C_{46}^{[44]} [W_4 W_4] + C_{46}^8 W_8 \\
\nonumber
W_4 W_8 & \sim C_{48}^4 W_4 + C_{48}^6 W_6 + C_{48}^8 W_8 + C_{48}^{[44]} [W_4 W_4] \\
\nonumber
& + C_{48}^{10} W_{10} + C_{48}^{[44]^{(2)}} [W_4 W_4]^{(2)} + C_{48}^{[46]} [W_4 W_6] + C_{48}^{[46]^{(1)}} [W_4 W_6]^{(1)} \\
\nonumber
W_6 W_6 & \sim C_{66}^0 \mathbbm{1} + C_{66}^4 W_4 + C_{66}^6 W_6 + C_{66}^8 W_8 + C_{66}^{[44]} [W_4 W_4] \\
& + C_{66}^{10} W_{10} + C_{66}^{[44]^{(2)}} [W_4 W_4]^{(2)} + C_{66}^{[46]} [W_4 W_6] + C_{66}^{[46]^{(1)}} [W_4 W_6]^{(1)} \\
\nonumber
W_4 W_{10} & \sim C_{4,10}^4 W_4 + C_{4,10}^6 W_6 + C_{4,10}^8 W_8 + C_{4,10}^{[44]} [W_4 W_4] \\
\nonumber
& + C_{4,10}^{10} W_{10} + C_{4,10}^{[44]^{(2)}} [W_4 W_4]^{(2)} + C_{4,10}^{[46]} [W_4 W_6] + C_{4,10}^{[46]^{(1)}} [W_4 W_6]^{(1)} \\
\nonumber
& + C_{4,10}^{12} W_{12} + C_{4,10}^{[48]} [W_4 W_8] + C_{4,10}^{[66]} [W_6 W_6] + C_{4,10}^{[444]} [W_4 W_4 W_4] \\
\nonumber
& + C_{4,10}^{[44]^{(4)}} [W_4 W_4]^{(4)} + C_{4,10}^{[46]^{(2)}} [W_4 W_6]^{(2)} + C_{4,10}^{[48]^{(1)}} [W_4 W_8]^{(1)} + C_{4,10}^{[46]^{(3)}} [W_4 W_6]^{(3)} \\
\nonumber
W_6 W_8 & \sim C_{68}^4 W_4 + C_{68}^6 W_6 + C_{68}^8 W_8 + C_{68}^{[44]} [W_4 W_4] \\
\nonumber
& + C_{68}^{10} W_{10} + C_{68}^{[44]^{(2)}} [W_4 W_4]^{(2)} + C_{68}^{[46]} [W_4 W_6] + C_{68}^{[46]^{(1)}} [W_4 W_6]^{(1)} \\
\nonumber
& + C_{68}^{12} W_{12} + C_{68}^{[48]} [W_4 W_8] + C_{68}^{[66]} [W_6 W_6] + C_{68}^{[444]} [W_4 W_4 W_4] \\
\nonumber
& + C_{68}^{[44]^{(4)}} [W_4 W_4]^{(4)} + C_{68}^{[46]^{(2)}} [W_4 W_6]^{(2)} + C_{68}^{[48]^{(1)}} [W_4 W_8]^{(1)} + C_{68}^{[46]^{(3)}} [W_4 W_6]^{(3)}
\end{align}
Here we only listed the primary fields that appear in the OPE. We used the brackets to denote the primary projection of the normal ordered product (possibly with derivatives). Assuming this ansatz for operator product expansions, the Jacobi identities impose many relations between the structure constants. The equations up to sum of spins $12$ are given in appendix \ref{primaryeqns}. Looking at these equations we observe the following: apart from the central charge $c$ there is essentially one undetermined structure constant which we choose to be the scale-invariant ratio
\begin{equation}
\label{parameterx}
x \equiv \frac{C_{46}^6}{C_{44}^4}.
\end{equation}
All other undetermined parameters can be chosen arbitrarily by rescaling the generating fields. We can choose $C_{44}^4$ to take any value by rescaling $W_4$. Afterwards the constant $C_{44}^6$ (assuming it to be generically non-vanishing) can be chosen to take any value if we appropriately rescale $W_6$. At dimension $8$ we have apart from $W_8$ a new composite primary $[W_4 W_4]$ so now we may freely choose $C_{46}^8$ and $C_{46}^{[44]}$ by redefinition of $W_8$ (i.e. $W_8$ can be shifted by any multiple of $[W_4 W_4]$) etc. The conjecture of \cite{Candu:2012ne} proven in \cite{Kanade:2018qut} is that this continues to hold in all higher OPEs, i.e. that there exists a two-parametric family of algebras parametrized by the central charge and parameter $x$.

Note that unlike in \cite{Candu:2012ne} where the independent structure constant was chosen to be
\begin{equation}
\frac{C_{44}^0}{\left(C_{44}^4\right)^2} = \frac{c(5c+22)}{72(c+24)} -\frac{7c(c-1)(5c+22)}{72(2c-1)(7c+68)} x + \frac{c(c-1)(c+24)(5c+22)}{12(2c-1)(7c+68)^2} x^2,
\end{equation}
here we will work instead with $x$ defined by (\ref{parameterx}). The advantage of working with this parameter is that we avoid the ambiguity of choosing a square root when solving for other structure constants. We will see that the group of duality symmetries of the algebra will have order $6$ just like the triality in $\mathcal{W}_\infty$. The $6$ other spurious solutions found in \cite{Candu:2012ne} are not symmetries of the algebra because they don't preserve the invariant ratio (\ref{parameterx}) and thus not all of the structure constants of even spin $\mathcal{W}_\infty$.

\subsection{Minimal representations}
In order to study the dualities of the algebra, we need to parametrize the algebra in terms of rank-like parameter. In \cite{Candu:2012ne} it was done by applying the method of \cite{Hornfeck:1993kp} and comparing the conformal dimensions of the minimal representations of the algebra obtained by solving the Jacobi identities with the Drinfe\v{l}d-Sokolov reduction applied to $B$, $C$ and $D$ series of simple Lie algebras. Minimal representations of even spin $\mathcal{W}_\infty$ are those representations whose character is (for generic values of parameters)
\begin{equation}
\chi_{min} = \frac{q^{h_m}}{1-q} \chi_{vac}.
\end{equation}
Following \cite{Candu:2012ne}, we assume OPE of minimal primary $\phi_m$ with $\mathcal{W}$-algebra generators to be of the form
\begin{align}
W_4 \phi_m & \sim w_{m4} \phi_m \\
W_6 \phi_m & \sim w_{m6} \phi_m + C_{6m}^{[4m]} [W_4 \phi_m] + C_{6m}^{[4m]^{(1)}} [W_4 \phi_m]^{(1)}
\end{align}
where the notation is like in the previous section. The Jacobi identity $(W_4 W_4 \phi_m)$ fixes the OPE and imposes the following constraints on the conformal dimension and higher spin charges:
\begin{align}
\nonumber
\frac{w_{m4}}{C_{44}^4} & = -\frac{h_m(ch_m+c+3h_m^2-7h_m+2)(2ch_m+c+16h_m^2-10h_m)}{12(2c^2h_m^2-2c^2h_m-9c^2+36ch_m^3-147ch_m^2+120ch_m-6c+24h_m^3+10h_m^2-28h_m)} \\
\frac{w_{m6} C_{44}^6}{(C_{44}^4)^2} & = \frac{(c-1)(5c+22)^2h_m(ch_m+c+3h_m^2-7h_m+2)}{54(c+24)(2c-1)(7c+68)} \times \\
\nonumber
& \times \frac{(ch_m+2c+15h_m^2-26h_m+8)(2ch_m+c+16h_m^2-10h_m)(2ch_m+3c+48h_m^2-28h_m)}{(2c^2h_m^2-2c^2h_m-9c^2+36ch_m^3-147ch_m^2+120ch_m-6c+24h_m^3+10h_m^2-28h_m)^2}
\end{align}
together with the following equation restricting $h_m$:
\begin{multline}
(4c^3h_m^2x-28c^3h_m^2-4c^3h_mx+28c^3h_m-18c^3x+91c^3+72c^2h_m^3x \\
-364c^2h_m^3-198c^2h_m^2x+988c^2h_m^2+144c^2h_mx-715c^2h_m-444c^2x+1010c^2 \\
+1776ch_m^3x-4040ch_m^3-7036ch_m^2x+12646ch_m^2+5704ch_mx-9616ch_m-288cx \\
+1224c+1152h_m^3x-4896h_m^3+480h_m^2x+3944h_m^2-1344h_mx-272h_m) \times \\
\times (12c^3h_m^2x-14c^3h_m^2-12c^3h_mx+14c^3h_m-54c^3x+168c^3+216c^2h_m^3x \\
-672c^2h_m^3-594c^2h_m^2x+3287c^2h_m^2+432c^2h_mx-2783c^2h_m-1332c^2x+1548c^2 \\
+5328ch_m^3x-6192ch_m^3-21108ch_m^2x+31544ch_m^2+17112ch_mx-26900ch_m-864cx \\
-816c+3456h_m^3x+3264h_m^3+1440h_m^2x-16592h_m^2-4032h_mx+14144h_m) = 0
\end{multline}
Given $h_m$ and $c$, there are thus two possible values for the structure constant $x$ of the algebra,
\begin{align}
\label{hmintox1}
x & = \frac{(2c-1)(7c+68)(h_m-4)(ch_m+3c+48h_m^2-52h_m)}{6(c+24)(2c^2h_m^2-2c^2h_m-9c^2+36ch_m^3-147ch_m^2+120ch_m-6c+24h_m^3+10h_m^2-28h_m)} \\
\label{hmintox2}
x & = \frac{(7c+68)(4c^2h_m^2-4c^2h_m-13c^2+52ch_m^3-180ch_m^2+141ch_m-18c+72h_m^3-58h_m^2+4h_m)}{2(c+24)(2c^2h_m^2-2c^2h_m-9c^2+36ch_m^3-147ch_m^2+120ch_m-6c+24h_m^3+10h_m^2-28h_m)}.
\end{align}
so knowing $c$ and $h_m$ still does not determine the algebra uniquely. Solving next the Jacobi identity $(W_4 W_6 \phi_m)$ determines the charge $w_{m8}$ but more importantly also picks the solution (\ref{hmintox1}) rather than (\ref{hmintox2}). This means that the structure constant $x$ of the algebra can be uniquely determined once we know the dimension of the minimal representation (and the central charge).

Given $c$ and $x$ there are generically three solutions of the equation for the conformal dimension of the minimal representation. This means that there are three minimal representations of even spin $\mathcal{W}_\infty$ and as we will see they are all permuted by the triality symmetry of the algebra. This is slightly different than in the case of $\mathcal{W}_\infty$ where we have six minimal representations, but there we have to use the triality symmetry together with the charge conjugation symmetry to find all of these representations. This is all consistent with the fact that while the minimal representations of $\mathcal{W}_\infty$ are charged (i.e. transform under charge conjugation symmetry which changes sign of odd spin fields), the even spin $\mathcal{W}_\infty$ has no conjugation symmetry.

\subsection{Parametrizations}

Analogously to \cite{Prochazka:2014gqa} we can introduce a redundant but triality covariant parametrization of the structure constants using three parameters permuted by the triality symmetry. Let us introduce three parameters $(\mu_1,\mu_2,\mu_3)$ such that the conformal dimensions of the minimal representations are
\begin{equation}
h_{m1} = \frac{1+\mu_1}{2}, \quad\quad h_{m2} = \frac{1+\mu_2}{2}, \quad\quad h_{m3} = \frac{1+\mu_3}{2}.
\end{equation}
These parameters are not independent but satisfy
\begin{equation}
\label{muconstraint}
\frac{1}{\mu_1} + \frac{1}{\mu_2} + \frac{1}{\mu_3} = 0
\end{equation}
just like in $\mathcal{W}_\infty$. The central charge in terms of these parameters is
\begin{equation}
c = \frac{(\mu_1+1)(\mu_2+1)(\mu_3+1)}{2}.
\end{equation}
We can also introduce a parameter $\psi$,
\begin{align}
\label{etapsi}
\mu_1 = \eta, \quad\quad \mu_2 = -\frac{\eta}{\psi}, \quad\quad \mu_3 = \frac{\eta}{\psi-1}.
\end{align}
The parameter $\psi$ is natural parameter from point of view of Drinfe\v{l}d-Sokolov reductions -- is the DS level shifted such that the critical level is at $\psi=0$. It also agrees with the Kapustin-Witten parameter $\Psi$ in Gaiotto-Rap\v{c}\'{a}k construction \cite{Gaiotto:2017euk}. Under triality transformations it transforms by fractional linear transformations permuting $(0,1,\infty)$. The other parameter $\eta$ measures the overall scale of $\mu_j$ (and is equal to one of $\mu_j$ depending on the triality frame). The condition (\ref{muconstraint}) is identically satisfied. The central charge takes a simple form
\begin{equation}
\label{etapsitoc}
c = \frac{(\eta+1)(\psi-\eta)(\eta+\psi-1)}{2(\psi-1)\psi}.
\end{equation}
There is one more parametrization of the algebra used in \cite{Kanade:2018qut}. The parameter $\lambda$ used there is related to $x$ by
\begin{equation}
x = \frac{(7c+68)(1+49\lambda-49\lambda c)}{84\lambda(1-c)(24+c)}.
\end{equation}
which is a fractional linear transformation so at given generic $c$ the correspondence between $x$ and $\lambda$ is one-to-one.

\subsection{Orthogonal cosets}
We may now identify the parameters of even spin $\mathcal{W}_\infty$ with parameters of orthogonal cosets expected to have even spin $\mathcal{W}_\infty$ symmetry. Using the formula for the central charge of affine Lie algebra
\begin{equation}
\frac{k \dim \mathfrak{g}}{k + h^\vee}
\end{equation}
where $k$ is the level of affine Lie algebra $\hat{\mathfrak{g}}$ and $h^\vee$ is the dual Coxeter number, we can calculate the central charge of the coset
\begin{equation}
\label{socoset}
\frac{\mathfrak{so}(n)_k \times \mathfrak{so}(n)_1}{\mathfrak{so}(n)_{k+1}}
\end{equation}
and we find
\begin{equation}
c = \frac{kn(2n+k-3)}{2(n+k-1)(n+k-2)}
\end{equation}
(which is uniform for both $B$ and $D$ series of cosets). The conformal dimension of the minimal representation is \cite{Candu:2012ne,Gaberdiel:2011nt}
\begin{equation}
h_2 = \frac{2n+k-3}{2(n+k-2)}
\end{equation}
for $(\Box,\Box;\bullet)$ and
\begin{equation}
h_3 = \frac{k}{2(n+k-1)}
\end{equation}
for $(\bullet,\Box;\Box)$. Expressing $x$ in terms of $c$ and one of $h_j$, we can find the dimensions of the other two minimal representations. The third minimal representation has conformal dimension
\begin{equation}
h_1 = \frac{n}{2}.
\end{equation}
This is interesting because in even orthogonal case this is exactly the dimension of the additional Pfaffian generator that we might add to the truncation of even spin $\mathcal{W}_\infty$.

To identify the even spin $\mathcal{W}_{\infty}$ corresponding to these cosets, we can use the central charge together with one of the minimal dimensions in formula (\ref{hmintox1}) and this determines the parameter $x$ uniquely. It is also possible to verify explicitly that (\ref{hmintox1}) is the correct branch (and not the one given by (\ref{hmintox2})) by direct evaluation of the $k \to \infty$ limit of these orthogonal cosets. In that case the coset simplifies to
\begin{equation}
\frac{\mathfrak{so}(n)_1}{\mathfrak{so}(n)}
\end{equation}
which is well-known to be realized by the singlet part of VOA of $n$ free fermions with OPE
\begin{equation}
\psi_j(z) \psi_k(w) \sim \frac{\delta_{jk}}{z-w}.
\end{equation}
The central charge of this is $c = \frac{n}{2}$ while the invariant ratio of structure constants is
\begin{equation}
x = \frac{C_{46}^6}{C_{44}^4} = \frac{49(n-8)(7n+136)}{6(n+48)(19n-68)}.
\end{equation}
This exactly agrees with the first branch (\ref{hmintox1}). The expression for the parameter $x$ in terms of $n$ and $k$ is thus
\begin{multline}
{\small
x = \frac{(n-8)(7k+6n-13)(7k+8n-8)(k^2+2kn-3k-n+2)}{6(k^2n+48k^2+2kn^2+93kn-144k+48n^2-144n+96)}} \times \\
\times \frac{(7k^2n+136k^2+14kn^2+251kn-408k+136n^2-408n+272)}{\splitfrac{(19k^4n-68k^4+76k^3n^2-386k^3n+408k^3+94k^2n^3-599k^2n^2+1267k^2n-860k^2+}{+36kn^4-252kn^3+857kn^2-1336kn+744k+24n^4-52n^3-124n^2+376n-224)}}.
\end{multline}
This completely determines the map from $(n,k)$ parameters to $(c,x)$. The parameters $\psi$ and $\eta$ are
\begin{equation}
\psi = 2-n-k, \quad\quad \eta = n-1.
\end{equation}
In fact, $\psi$ is determined only up to a $S_3$ subgroup of M\"{o}bius transformations permuting $(0,1,\infty)$. The other five choices of $\psi$ correspond to 5 other embeddings related by the triality symmetry. In terms of parameters $(n,k)$, the following $6$ values correspond to the same even spin $\mathcal{W}_\infty$:
\begin{multline}
(n,k), \quad (n, 3-2n-k), \quad \left( \frac{k}{n+k-1}, \frac{n}{n+k-1} \right), \\
\left( \frac{k}{n+k-1}, \frac{2n+k-3}{n+k-1} \right), \quad \left(\frac{2n+k-3}{n+k-2}, \frac{k}{n+k-2} \right), \quad \left( \frac{2n+k-3}{n+k-2}, -\frac{n}{n+k-2} \right).
\end{multline}

\paragraph{Symplectic quotients}
We can use the duality between orthogonal and symplectic algebras to study the corresponding symplectic quotients. In general, the Grassmannian coset of the type
\begin{equation}
\frac{\mathfrak{so}(n)_k \times \mathfrak{so}(n)_l}{\mathfrak{so}(n)_{k+l}} \simeq \frac{\mathfrak{so}(k+l)_n}{\mathfrak{so}(k)_n \times \mathfrak{so}(l)_n}
\end{equation}
with central charge
\begin{equation}
\frac{kln(n-1)(2n+k+l-4)}{(n+k-2)(n+l-2)(n+k+l-2)}
\end{equation}
has a triality symmetry if we define three parameters $k_1=k, k_2=l, k_3=4-2n-k-l$ just like the unitary cosets. The unitary cosets have also a $\mathbbm{Z}_2$ symmetry which changes signs of all $k_j$ parameters. In the case of orthogonal cosets this $\mathbbm{Z}_2$ symmetry instead maps the cosets to symplectic ones\footnote{The reason that we have half-integer levels is a consequence of the usual convention for normalization of Killing form such that the length squared of long roots is $2$. In the $C_n$ case that we are considering this leads to dual Coxeter number $n+1$ which is half of we would get if we worked in more symmetric conventions where the length squared of short roots in $C_n$ would be $2$.}, i.e.
\begin{equation}
\frac{\mathfrak{sp}(2n)_{\frac{k}{2}} \times \mathfrak{sp}(2n)_{\frac{l}{2}}}{\mathfrak{sp}(2n)_{\frac{k+l}{2}}} \simeq \frac{\mathfrak{so}(-2n)_{-k} \times \mathfrak{so}(-2n)_{-l}}{\mathfrak{so}(-2n)_{-k-l}}.
\end{equation}
This means that the symplectic analogue of the cosets (\ref{socoset}) are cosets
\begin{equation}
\frac{\mathfrak{sp}(2n)_k \times \mathfrak{sp}(2n)_{-\frac{1}{2}}}{\mathfrak{sp}(2n)_{k-\frac{1}{2}}}
\end{equation}
of the central charge
\begin{equation}
-\frac{kn(4n+2k+3)}{(n+k+1)(2n+2k+1)}.
\end{equation}
The dimensions of the minimal representations are
\begin{equation}
h_1 = -n, \quad\quad h_2 = \frac{4n+2k+3}{4(n+k+1)}, \quad\quad h_3 = \frac{k}{2n+2k+1}.
\end{equation}
The simplest level $-\frac{1}{2}$ representation may be realized as singlet part of VOA of $2n$ free symplectic bosons with OPE
\begin{equation}
\xi_j(z) \xi_k(w) \sim \frac{\omega_{jk}}{z-w}
\end{equation}
where $\omega_{jk}$ is a non-degenerate symplectic form. The structure constants of this symplectic quotient algebra exactly agree with those of $2n$ free fermions (i.e. level $1$ even orthonormal coset) if we change the sign of $n$ everywhere.

\subsection{Drinfe\v{l}d-Sokolov reductions}
Let us summarize the central charges and dimensions of minimal representations in principal Drinfe\v{l}d-Sokolov reduction of $B_n$, $C_n$ and $D_n$ type algebras following \cite{Bouwknegt:1992wg,Candu:2012ne}. The first main formula that we will use is the expression for the central charge
\begin{equation}
\label{dscentralcharge}
c = \ell - 12 \big|\alpha_+ \rho + \alpha_- \rho^\vee \big|^2
\end{equation}
where $\ell$ is the rank of the Lie algebra, $\rho$ is the Weyl vector and $\rho^\vee$ the dual Weyl vector. The parameters $\alpha_{\pm}$ are defined as
\begin{equation}
\alpha_+ = \frac{1}{\sqrt{k+h^\vee}}, \quad\quad \alpha_- = -\sqrt{k+h^\vee}
\end{equation}
where $k$ is the level of the affine Lie algebra entering the Drinfe\v{l}d-Sokolov reduction and $h^\vee$ is the dual Coxeter number. The second useful formula is the formula for the dimension of maximally degenerate representation parametrized by the pair of highest (co)weights $(\Lambda^+,\Lambda^-)$ where $\Lambda^+$ is integral dominant weight and $\Lambda^-$ integral dominant co-weight:
\begin{equation}
\label{dsdimension}
h = \frac{1}{2} \langle\alpha_+ \Lambda_+ + \alpha_- \Lambda_-, \alpha_+ (\Lambda_++2\rho) + \alpha_- (\Lambda_-+2\rho^\vee) \rangle.
\end{equation}

\paragraph{Odd orthogonal case - $\mathfrak{so}(2n_B+1)$}
Turning now to Lie algebra $B_{n_B}$, from (\ref{dscentralcharge}) we find the central charge
\begin{equation}
c_B = -\frac{n_B(4n_B^2+2n_Bk_B+2k_B-3)(4n_B^2+2n_Bk_B-2n_B+k_B)}{2n_B+k_B-1}
\end{equation}
and from (\ref{dsdimension}) the minimal weights
\begin{equation}
h_{B2} = -\frac{n_B(2n_B+k_B-2)}{2n_B+k_B-1}
\end{equation}
(corresponding to $\Lambda_+ = \omega_1$) and
\begin{equation}
h_{B1} = \frac{1}{2} \left( 4n_B^2+2n_Bk_B-2n_B+k_B \right).
\end{equation}
(corresponding to $\Lambda_- = \omega_1^\vee$ which here agrees with $\omega_1$). These two weights are compatible with third minimal weight
\begin{equation}
h_{B3} = \frac{4n_B^2+2n_Bk_B+2k_B-3}{2(2n_B+k_B-2)}.
\end{equation}
The shifted level $\psi_B$ is
\begin{equation}
\psi_B = 2n_B-1+k_B
\end{equation}
and the scale parameter $\eta_B$ is
\begin{equation}
\eta_B = 2n_B \psi_B - 2n_B + \psi_B.
\end{equation}

\paragraph{Symplectic case - $\mathfrak{sp}(2n_C)$}
For Lie algebra $C_{n_C}$ we find the central charge
\begin{equation}
c_C = -\frac{n_C(2n_C^2+2n_Ck_C+2n_C+k_C)(4n_C^2+4n_Ck_C-2k_C-3)}{n_C+k_C+1}
\end{equation}
and dimensions of minimal representations
\begin{equation}
h_{C2} = -\frac{4n_C^2+4n_Ck_C-2k_C-3}{4(n_C+k_C+1)}
\end{equation}
(corresponding to $\Lambda_+ = \omega_1$) and
\begin{equation}
h_{C1} = n_C(2n_C+2k_C+1).
\end{equation}
(corresponding to $\Lambda_- = \omega_1^\vee$ which is twice as long as $\omega_1$). The third minimal weight compatible with these is
\begin{equation}
h_{C3} = \frac{2n_C^2+2n_Ck_C+2n_C+k_C}{2n_C+2k_C+1}.
\end{equation}
The shifted level $\psi_C$ is
\begin{equation}
\psi_C = 2n_C+2+2k_C
\end{equation}
and the scale parameter $\eta_C$
\begin{equation}
\eta_C = 2n_C \psi_C - 2n_C -1.
\end{equation}

\paragraph{Even orthogonal case - $\mathfrak{so}(2n_D)$}
In the case of $D_{n_D}$ the calculation is slightly simpler because the Lie algebra is simply laced. We find
\begin{equation}
c_D = -\frac{n_D(4n_D^2+2n_Dk_D-10n_D-2k_D+5)(4n_D^2+2n_Dk_D-8n_D-k_D+4)}{2n_D+k_D-2}
\end{equation}
and the minimal dimensions are
\begin{equation}
h_{D2} = -\frac{4n_D^2+2n_Dk_D-10n_D-2k_D+5}{4n_D+2k_D-4}
\end{equation}
(for $\Lambda_+=\omega_1$) and
\begin{equation}
h_{D1} = \frac{1}{2} \left( 4n_D^2+2n_Dk_D-8n_D-k_D+4 \right)
\end{equation}
(for $\Lambda_-=\omega_1$ since now the weights and co-weights agree). The third minimal weight compatible with these is simply
\begin{equation}
h_{D3} = n_D
\end{equation}
(just like in the case of orthogonal coset, this is compatible with assumption that the Pfaffian generating field transforms in the minimal representation of the algebra). Finally we define the shifted level $\psi_D$ to be
\begin{equation}
\psi_D = 2n_D-2+k_D
\end{equation}
and the parameter $\eta_D$ is
\begin{equation}
\eta_D = 2n_D \psi_D - 2n_D - \psi_D +1 = (2n_D-1)(\psi_D-1).
\end{equation}

\paragraph{Note on Pfaffian generator} Let us briefly discuss the Pfaffian generator of dimension $n_D$. There is no corresponding field in even spin $\mathcal{W}_\infty$, although we have just seen that one of the minimal primaries has exactly the correct conformal dimension. The reason for it is that it is unstable as we vary $n$. In this sense the $\mathcal{W}$-algebra of type $WD$ is not a truncation of even spin $\mathcal{W}_\infty$, only its $\mathbbm{Z}_2$ projection which removes the Pfaffian generator is \cite{Kanade:2018qut}. On the other hand, we will see in the next section when we discuss the Miura transformation that the Pfaffian generator can be naturally embedded into $\mathfrak{u}(1) \times \mathcal{W}_N$ truncation of $\mathcal{W}_{1+\infty}$ and actually this operator plays a crucial role in construction of the embedding of even spin $\mathcal{W}_\infty$ into $\mathcal{W}_{1+\infty}$.

\subsection{Gaiotto-Rap\v{c}\'{a}k}
\label{secgr}
In \cite{Gaiotto:2017euk} the authors found an interesting realization of $\mathcal{W}$-algebras in gauge theory setting. The theory they considered was four-dimensional twisted $\mathcal{N}=4$ super Yang-Mills theory with three semi-infinite co-dimension one defects meeting at co-dimension two subspace. The degrees of freedom living at this co-dimension $2$ subspace were found to be organized by a certain truncation of $\mathcal{W}_{1+\infty}$ algebra determined by the ranks of the gauge groups in three subsectors of the full four-dimensional space cut out by the co-dimension $1$ defects \cite{Gaiotto:2017euk,Prochazka:2017qum}. This setup can be modified by introducing an orientifold plane. The unitary gauge groups are then projected to orthosymplectic groups and one expects the degrees of freedom at co-dimension $2$ subspace to be reduced to even spin $\mathcal{W}_\infty$. Here we verify that the central charge formula derived in \cite{Gaiotto:2017euk} is compatible with the form of the central charge in even spin $\mathcal{W}_\infty$ and later that the orthosymplectic $Y$-algebras can be identified with the truncations of even spin $\mathcal{W}_\infty$.

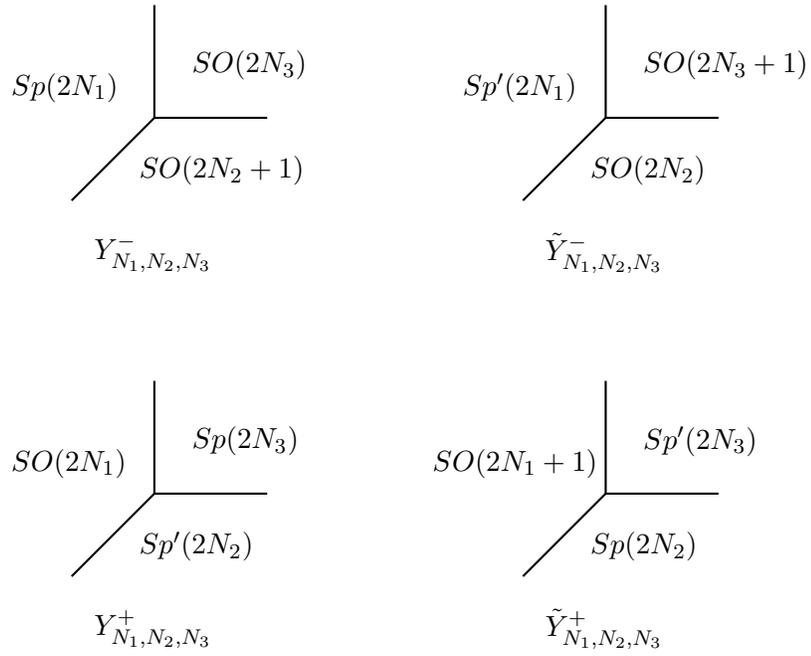
\begin{figure}
\begin{center}
\begin{tikzpicture}
\draw[thick] (0,5+0) -- (1.5,5+0);
\draw[thick] (0,5+0) -- (0,5+1.5);
\draw[thick] (0,5+0) -- (-1.1,5-1.1);
\node[text width=3cm] at (-0.4,5+0.4) {$Sp(2N_1)$};
\node[text width=3cm] at (1.3,5-0.7) {$SO(2N_2+1)$};
\node[text width=3cm] at (2,5+0.7) {$SO(2N_3)$};
\node[text width=3cm] at (0.7,5-1.8) {$Y^-_{N_1,N_2,N_3}$};
\draw[thick] (6,5+0) -- (6+1.5,5+0);
\draw[thick] (6,5+0) -- (6,5+1.5);
\draw[thick] (6,5+0) -- (6-1.1,5-1.1);
\node[text width=3cm] at (6-0.4,5+0.4) {$Sp^\prime(2N_1)$};
\node[text width=3cm] at (6+1.3,5-0.7) {$SO(2N_2)$};
\node[text width=3cm] at (6+2,5+0.7) {$SO(2N_3+1)$};
\node[text width=3cm] at (6+0.7,5-1.8) {$\tilde{Y}^-_{N_1,N_2,N_3}$};
\draw[thick] (0,0) -- (1.5,0);
\draw[thick] (0,0) -- (0,1.5);
\draw[thick] (0,0) -- (-1.1,-1.1);
\node[text width=3cm] at (-0.4,0.4) {$SO(2N_1)$};
\node[text width=3cm] at (1.3,-0.7) {$Sp^\prime(2N_2)$};
\node[text width=3cm] at (2,0.7) {$Sp(2N_3)$};
\node[text width=3cm] at (0.7,-1.8) {$Y^+_{N_1,N_2,N_3}$};
\draw[thick] (6,0) -- (6+1.5,0);
\draw[thick] (6,0) -- (6,1.5);
\draw[thick] (6,0) -- (6-1.1,-1.1);
\node[text width=3cm] at (6-0.8,0.4) {$SO(2N_1+1)$};
\node[text width=3cm] at (6+1.3,-0.7) {$Sp(2N_2)$};
\node[text width=3cm] at (6+2,0.7) {$Sp^\prime(2N_3)$};
\node[text width=3cm] at (6+0.7,-1.8) {$\tilde{Y}^+_{N_1,N_2,N_3}$};
\end{tikzpicture}
\label{yalgebrafig}
\caption{Gaiotto-Rap\v{c}\'{a}k orthosymplectic $Y$-algebras}
\end{center}
\end{figure}

As discussed in \cite{Gaiotto:2017euk} there are actually four different ways how to introduce an orientifold plane in the theory leading to four different families of $Y$-algebras. They are shown in figure \ref{yalgebrafig}. Although we expect that the orthosymplectic algebras constructed in \cite{Gaiotto:2017euk} should be truncations of even spin $\mathcal{W}_\infty$, to identify the parameters one would need to know the central charge and one of the structure constants. Unfortunately only the central charge was calculated in \cite{Gaiotto:2017euk}. On the other hand, the orthosymplectic $Y$-algebras transform nicely under triality transformations and the Kapustin-Witten parameter $\Psi$ has exactly the properties of the parameter $\psi$ introduced in (\ref{etapsi}) so one can try to identify these $\Psi$ with $\psi$. Finding rational expressions for minimal dimensions and compatibility with various truncations and restrictions would already be a big hint of correctness of the proposed identification.

\paragraph{Algebra $Y^-_{N_1,N_2,N_3}$}
Starting with the first algebra of the figure \ref{yalgebrafig}, $Y^-$, the central charge is given by (\ref{etapsitoc}) with \footnote{Some of the formulas in \cite{Gaiotto:2017euk} contain typos. I would like to thank to Miroslav Rap\v{c}\'{a}k for sharing with me the corrected expressions for these.}
\begin{equation}
\eta^- = 1 + 2(N_1-N_3) - (1 + 2(N_2-N_3)) \psi.
\end{equation}
From this we can immediately find the three $\mu$ parameters of the algebra using (\ref{etapsi}). The next algebra is $\tilde{Y}^-$. The central charge calculated in \cite{Gaiotto:2017euk} is of the form (\ref{etapsitoc}) with
\begin{equation}
\tilde{\eta}^- = 2(N_1-N_3) + (1 - 2(N_2-N_3)) \psi.
\end{equation}
The third algebra, $Y^+$ has parameter $\eta$ equal to
\begin{equation}
\eta^+ = -1 + 2(N_1-N_3) - 2(N_2-N_3)\psi.
\end{equation}
The last algebra of figure \ref{yalgebrafig} is $\tilde{Y}^+$ with $\eta$ parameter equal to
\begin{equation}
\tilde{\eta}^+ = 2(N_1-N_3) - 2(N_2-N_3)\psi.
\end{equation}
In all four cases we get nice polynomial expressions for $\eta$ which has the same structure as for cosets of Drinfe\v{l}d-Sokolov reductions. Let's summarize some of the properties of these algebras have:
\begin{enumerate}
\item The parameters of even spin $\mathcal{W}_\infty$ don't change if we shift all three $N_j$ parameters at the same time by a constant. This is analogous to what happens in $\mathcal{W}_\infty$ and is a consequence of (\ref{muconstraint}). This doesn't mean though that the $Y_{N_1,N_2,N_3}$ algebras are the same: only their simple quotient is expected to be the same. In the case of $\mathcal{W}_\infty$ this is discussed in \cite{Prochazka:2017qum} and in particular in \cite{Prochazka:2018tlo} in connection with free field representations.
\item The transformation $\psi \leftrightarrow 1-\psi$ in parametrization (\ref{etapsi}) exchanges $\mu_1 \leftrightarrow \mu_3$ and transforms $\eta$ only by its action on $\psi$. The effect on orthosymplectic $Y$-algebras is
\begin{align}
\nonumber
Y^-(N_1,N_2,N_3) \leftrightarrow \tilde{Y}^-(N_1,N_3,N_2), \quad & \quad Y^+(N_1,N_2,N_3) \leftrightarrow Y^+(N_1,N_3,N_2) \\
\tilde{Y}^+(N_1,N_2,N_3) & \leftrightarrow \tilde{Y}^+(N_1,N_3,N_2)
\end{align}
which is exactly the claim in \cite{Gaiotto:2017euk}. Note that pictorially it exchanges the upper right and lower right gauge groups in figure \ref{yalgebrafig} from where the action on ranks and type of $Y$-algebra is obvious.
\item To see the effect of the transformation $\psi \to \frac{1}{\psi}$ on $Y$-algebras it's better to work directly with $\mu_j$ parameters. We find
\begin{align}
\nonumber
Y^-(N_1,N_2,N_3) \leftrightarrow Y^-(N_2,N_1,N_3), \quad & \quad \tilde{Y}^-(N_1,N_2,N_3) \leftrightarrow Y^+(N_2,N_1,N_3) \\
\tilde{Y}^+(N_1,N_2,N_3) & \leftrightarrow \tilde{Y}^+(N_2,N_1,N_3)
\end{align}
again in agreement with \cite{Gaiotto:2017euk}.
\item The third operation of exchanging two gauge groups corresponds to $\psi \to \frac{\psi}{\psi-1}$. The action on $Y$-algebras is
\begin{align}
\nonumber
Y^-(N_1,N_2,N_3) \leftrightarrow Y^+(N_3,N_2,N_1), \quad & \quad \tilde{Y}^-(N_1,N_2,N_3) \leftrightarrow \tilde{Y}^-(N_3,N_2,N_1) \\
\tilde{Y}^+(N_1,N_2,N_3) & \leftrightarrow \tilde{Y}^+(N_3,N_2,N_1).
\end{align}
The fact that $Y^-$ and $Y^+$ exchange their roles is again manifest in figure \ref{yalgebrafig}.
\item At the level of the parameters of the universal algebra, all four orthosymplectic $Y$-algebras are connected formally by half-integer shifts of rank parameters: apart from the shift
\begin{equation}
\tilde{\eta}^+(N_1,N_2,N_3) = \eta^+\left(N_1+\tfrac{1}{2}, N_2, N_3\right)
\end{equation}
used already in \cite{Gaiotto:2017euk} and its generalization
\begin{equation}
\tilde{\eta}^-(N_1,N_2,N_3) = \eta^-\left(N_1,N_2-\tfrac{1}{2},N_3+\tfrac{1}{2}\right)
\end{equation}
which also involves the transformation $\psi \to 1-\psi$ we have also formally
\begin{align}
\nonumber
\tilde{\eta}^-(N_1,N_2,N_3) & = \eta^-\left(N_1-\tfrac{1}{2}, N_2-1, N_3\right) \\
\eta^+(N_1,N_2,N_3) & = \eta^-\left(N_1-1, N_2-\tfrac{1}{2}, N_3\right) \\
\nonumber
\tilde{\eta}^+(N_1,N_2,N_3) & = \eta^-\left(N_1-\tfrac{1}{2}, N_2-\tfrac{1}{2}, N_3\right)
\end{align}
which allows to map the parameters of any two orthosymplectic $Y$-algebras thought of in terms of the universal even spin algebra.
\end{enumerate}

\section{Miura transformation and quadratic basis}
In this section we show how we can use the results of \cite{Lukyanov:1989gg} to find a free field representation of even spin $\mathcal{W}_\infty$ and embed it in $\mathcal{W}_{1+\infty}$. First of all, recall that given $N$ free fields with currents satisfying OPE
\begin{equation}
J_j(z) J_k(w) \sim \frac{\delta_{jk}}{(z-w)^2}
\end{equation}
we can construct Miura operator
\begin{equation}
(\alpha_0 \partial + J_1(z)) \cdots (\alpha_0 \partial + J_N(z)) = \sum_{k=0}^N U_k(z) (\alpha_0 \partial)^{N-k}
\end{equation}
and the currents $U_k(z)$ defined in this way represent algebra $\widehat{u(1)} \times \mathcal{W}_N$ \cite{Fateev:1987zh,luk1988quantization} and moreover the operator product expansions are quadratic in this basis \cite{luk1988quantization,Prochazka:2014gqa}.

An important observation of \cite{Lukyanov:1989gg} is that the fields appearing in the OPE of the generating field of the highest spin $W_N$ in $\widehat{u(1)} \times \mathcal{W}_N$ generate an even spin subalgebra. Following \cite{Bouwknegt:1992wg}, we can define fields $V_j(z)$ by
\begin{equation}
\label{ununbs}
U_N(z) U_N(w) = \frac{a(N-1)}{(z-w)^{2N}} + \sum_{k=1}^{N-1} \frac{(-1)^k a(N-1-k)}{(z-w)^{2N-2k}} \left[ V_{2k}(z) + V_{2k}(w) \right]
\end{equation}
where we choose the normalization factors as
\begin{equation}
a(j) = \prod_{r=1}^j \left(1-(2j)(2j+1)\alpha_0^2\right).
\end{equation}
These are not so easy to calculate explicitly at larger values of $N$, because even if we are interested in fields $V_{2j}$ with $j$ small, we still need to know the OPE of $U_N$ with itself. Fortunately, we can use the result that the OPE can be written in the form \cite{Prochazka:2014gqa}
\begin{equation}
\label{bilocnn}
U_N(z) U_N(w) = \sum_{l+m \leq 2N} C_{NN}^{lm}(\alpha_0,N) \frac{U_{lm}(z,w)}{(z-w)^{2N-l-m}}
\end{equation}
where $U_{lm}(z,w)$ are certain bi-local fields of the form $(U_l U_m)(w) + derivatives$ and can be explicitly written in terms of fields $U_j(z) U_k(w)$ with $j+k \leq l+m$. More concretely they are equal to
\begin{equation}
U_{lm}(z,w) = \sum_{j+k \leq l+m} \frac{D_{lm}^{jk} U_j(z) U_k(w)}{(z-w)^{l+m-j-k}}
\end{equation}
and the matrix of constants $D_{lm}^{jk}$ is the inverse of the matrix of structure constants $C_{lm}^{jk}$ (considering $(j,k)$ and $(l,m)$ as bi-indices as explained in \cite{Prochazka:2014gqa}). The structure constants for OPE of $U_N$ with itself in our situation simplify to
\begin{equation}
C_{NN}^{jk}(\alpha_0,N) = (-1)^{\frac{j-k}{2}} \prod_{r=1}^{\frac{2N-j-k-2}{2}} \left( 1-2r(2r+1)\alpha_0^2 \right) = (-1)^{\frac{j-k}{2}} a\left(\frac{2N-j-k-2}{2}\right)
\end{equation}
for $j+k$ even and to
\begin{align}
\nonumber
C_{NN}^{jk}(\alpha_0,N) & = (-1)^{\frac{j-k-1}{2}} (2n-j-k-1) \alpha_0 \prod_{r=1}^{\frac{2N-j-k-3}{2}} \left( 1-2r(2r+1)\alpha_0^2 \right) \\
& = (-1)^{\frac{j-k-1}{2}} (2n-j-k-1) \alpha_0 a\left(\frac{2N-j-k-3}{2}\right)
\end{align}
for $j+k$ odd. In both cases, these depend only on the sum $j+k$ (except for an overall sign). Now the problem with extracting lower spin fields $V_{2j}$ at larger values of $N$ is solved, because we can use the expression (\ref{bilocnn}) to directly extract $V_{2s}$ fields, a calculation which involves knowledge of OPE of fields of spin $\leq 2s$ only.

Let use write a formula that we can use to extract the generators of even spin $\mathcal{W}_\infty$ in terms of those of $\mathcal{W}_{1+\infty}$. For that, we Taylor expand (\ref{ununbs}) at $z=w$ obtaining an ordinary OPE. The coefficient of pole of order $2N-2s$ is
\begin{equation}
2(-1)^s a(N-1-s) V_{2s}(w) + \sum_{r=1}^{s-1} \frac{(-1)^r a(N-1-r)}{(2(s-r))!} V_{2r}^{(2s-2r)}(w).
\end{equation}
On the other hand, the coefficient of the same pole in the expansion of the form (\ref{bilocnn}) is equal to \footnote{Note that the multiplication of the matrix components with the components of the inverse matrix does not give the identity matrix because of the restriction on the range of the indices.}
\begin{equation}
\sum_{l+m \leq j+k \leq 2s} C_{NN}^{jk} D_{jk}^{lm} \, \mathtt{OPEPole}[l+m-2s][U_l,U_m]
\end{equation}
which is a convenient expression involving only OPE of fields with spins $\leq 2s$. Equating these last two expressions, we can recurrently calculate expressions for $V_{2s}$ fields in terms of $U_j$ fields. The first few fields are given in the next section. It is a non-trivial check of our calculations that the OPEs of $V_j$ fields close.

\subsection{Map between parameters of even spin $\mathcal{W}_\infty$ and $\mathcal{W}_{1+\infty}$}
Using the OPE of the Pfaffian field as described in the previous section we can extract first two $V_j$ fields:
\begin{align}
\nonumber
V_2 & = U_2 - \frac{1}{2} (U_1 U_1) - (N-1)\alpha_0 U_1^\prime \\
\nonumber
V_4 & = U_4 - (U_1 U_3) + \frac{1}{2} (U_2 U_2) - (N-2)\alpha_0 U_3^\prime + (N-2)\alpha_0 (U_1 U_2^\prime) - (N-2)\alpha_0 (U_1^\prime U_2) \\
& + \frac{1}{4} U_2^{\prime\prime} - \frac{N-1}{4} (U_1^{\prime\prime} U_1) + \frac{4N^2 \alpha_0^2 - 14N \alpha_0^2 + 12\alpha_0^2 - 1}{4} (U_1^\prime U_1^\prime) \\
\nonumber
& + \frac{(N-1)\alpha_0(4N^2\alpha_0^2-14N\alpha_0^2+12\alpha_0^2-N-1)}{12} U_1^{\prime\prime\prime}.
\end{align}
The third field, $V_6$, is given in the appendix. Since $V_2$ and $V_4$ generate even spin $\mathcal{W}_\infty$ subalgebra, we can identify the parameters of even spin $\mathcal{W}_\infty$ in terms of those of $\mathcal{W}_{1+\infty}$ ($N$ and $\alpha_0$). We first need to find the stress-energy tensor which is simply $-V_2$ and the primary combination of spin $4$ fields and spin $6$ fields to extract the central charge and the parameter $x$. The result is
\begin{equation}
c = N(1-2(N-1)(2N-1)\alpha_0^2)
\end{equation}
and
\begin{multline}
x = \Big[ (N-4)(-6\alpha_0^2+4\alpha_0^2N^2+10\alpha_0^2N-49)(28\alpha_0^2N^3-42\alpha_0^2N^2+14\alpha_0^2N-7N-68) \times \\
\times (4\alpha_0^2N^2-4\alpha_0^2N-1) \Big] \Big/ \Big[ 6(4\alpha_0^2N^3-6\alpha_0^2N^2+2\alpha_0^2N-N-24) \times \\
\times (12\alpha_0^2+16\alpha_0^4N^5-40\alpha_0^4N^4-40\alpha_0^4N^3-80\alpha_0^2N^3+100\alpha_0^4N^2 + \\
+ 302\alpha_0^2N^2-36\alpha_0^4N-204\alpha_0^2N+19N-34) \Big]
\end{multline}
Expressing $N$ and $\alpha_0^2$ in terms of parameters $\lambda$ of $\mathcal{W}_{\infty}$ \cite{Prochazka:2014gqa},
\begin{equation}
c_{\infty} = (\lambda_1-1)(\lambda_2-1)(\lambda_3-1)
\end{equation}
and
\begin{equation}
\frac{1}{\lambda_1} + \frac{1}{\lambda_2} + \frac{1}{\lambda_3} = 0
\end{equation}
and $\lambda_3 = N$ we can write even spin $\mathcal{W}_\infty$ parameters as
\begin{align}
\nonumber
h_1 & = 1 + \lambda_1 + \frac{\lambda_1}{2\lambda_2} = \frac{1+\mu_1}{2} \\
h_2 & = 1 + \lambda_2 + \frac{\lambda_1}{2\lambda_2} = \frac{1+\mu_2}{2} \\
\nonumber
h_3 & = \lambda_3 = \frac{1+\mu_3}{2}.
\end{align}
In terms of parameters $\mu$ we have
\begin{align}
\nonumber
\mu_1 & = \frac{\lambda_1+\lambda_2+2\lambda_1\lambda_2}{\lambda_2} \\
\mu_2 & = \frac{\lambda_1+\lambda_2+2\lambda_1\lambda_2}{\lambda_1} \\
\nonumber
\mu_3 & = - \frac{\lambda_1+\lambda_2+2\lambda_1\lambda_2}{\lambda_1+\lambda_2}.
\end{align}
For reference, the map between parameters $(N,\alpha_0)$ and $\mu_j$ is
\begin{align}
N & = \frac{\mu_1+\mu_2-\mu_1\mu_2}{2(\mu_1+\mu_2)} = \frac{\mu_3}{2} \left( 1-\frac{1}{\mu_1}-\frac{1}{\mu_2} \right) \\
\alpha_0^2 & = -\frac{(\mu_1+\mu_2)^2}{\mu_1 \mu_2} = -\frac{\mu_1 \mu_2}{\mu_3^2} = \frac{(1-\psi)^2}{\psi}.
\end{align}
We see that the embedding of even spin $\mathcal{W}_\infty$ in $\mathcal{W}_{1+\infty}$ breaks the triality symmetry of $\mathcal{W}_{1+\infty}$ to a $\mathbbm{Z}_2$ exchanging $\lambda_1 \leftrightarrow \lambda_2$ or $\mu_1 \leftrightarrow \mu_2$. This is related to the fact that the Miura transformation depends on a choice of a preferred direction. So although both algebras have the triality symmetry, the triality in $\mathcal{W}_{1+\infty}$ does not restrict to triality in even spin $\mathcal{W}_\infty$. The choice of even spin $\mathcal{W}_\infty$ subalgebra in $\mathcal{W}_{1+\infty}$ breaks the triality symmetry to $\mathbbm{Z}_2$ but when restricted to this subalgebra, the duality is enhanced to a triality of the subalgebra. This is analogous to enhancement of duality to triality in unitary Grassmannian cosets when one of the levels is one. We also see that there are at least six ways of embedding even spin $\mathcal{W}_\infty$ in $\mathcal{W}_{1+\infty}$, each associated to different asymptotic direction (times two because of the complex conjugation in $\mathcal{W}_\infty$).

\subsection{Operator product expansions in $V_j$ basis}
As a result of our definition of $V_j$ fields they are quadratic composites of the $U_j$ fields. Since $\mathcal{W}_{1+\infty}$ is filtered with degree given by the number of $U_j$ fields in each term and since the operator product expansions preserve the degree, we can also expect the operator product expansions of $V_j$ field to satisfy quadratic operator product expansions.

To fix these, we can first calculate the OPE of $V_2$ with $V_{2s}$,
\begin{align}
\nonumber
V_2(z) V_{2j}(w) & \sim \frac{4N (-1)^j \left[2^{2j+2}-1\right] a(N-1) B_{2j+2}}{(2j+2)a(N-j-1)} \frac{\mathbbm{1}}{(z-w)^{2j+2}} \\
\nonumber
& + \sum_{k=1}^{j-1} \frac{8(N-k) (-1)^{j-k} \left[2^{2j-2k+2}-1\right] a(N-k-1) B_{2j-2k+2}}{(2j-2k+2) a(N-j-1)} \frac{V_{2k}(w)}{(z-w)^{2j-2k+2}} \\
& + \frac{(derivatives)}{(z-w)^{\geq 3}} - \frac{2j V_{2j}(w)}{(z-w)^2} - \frac{\partial V_{2j}(w)}{z-w}.
\end{align}
where $B_n$ are the Benoulli numbers. The Jacobi identity $(V_2 V_2 V_{2j})$ fixes all the derivative terms in $V_2 V_{2j}$ OPE. The OPE of $V_4$ with itself is given in the appendix \ref{apquad}. With this input (actually only the coefficient of the identity and of $V_4$ in $V_4 V_4$ OPE is necessary) the Jacobi identities determine all the other operator product expansions. The resulting OPEs have the following properties
\begin{enumerate}
\item the operator product expansions are purely quadratic, i.e. all the operators appearing in the singular part of the OPE are normal ordered products of (at most) two $V_j$ fields and their derivatives. This is analogous to the case of $\mathcal{W}_{1+\infty}$ \cite{luk1988quantization,Prochazka:2014gqa}.
\item All the structure constants are polynomial functions of $N$ and $\alpha_0$. This is again analogous to \cite{Prochazka:2014gqa,Eberhardt:2019xmf}.
\item Unlike in $\mathcal{W}_{1+\infty}$, the derivatives do not seem to be simply summable into bi-local fields (this seems to be the case even after a simple linear redefinition of the fields). This is probably related to different form of the Miura operator which is `folded'. As a consequence of this, the calculation of commutation relations between mode operators is more involved because one needs to consider terms with derivatives.
\item We verified these claims for OPE of fields $V_j$ and $V_k$ with $j+k \leq 20$. At every step determination of OPE reduces to solution of linear equations for the coefficients of the quadratic composites in the OPE.
\end{enumerate}

For later purposes, it is useful to determine at each even spin the primary field $W_{2j}$ whose pole of order $4j$ with all dimension $2j$ fields not involving $V_{2j}$ vanishes, i.e. field which is orthogonal to all lower dimension fields and their derivatives and composites. Actually we don't even need to require this to be primary, it is a consequence of being orthogonal to lower composites. The special property of this field is that it is the field whose two-point function vanishes for truncations of the algebra. We can thus avoid searching for zeros of Kac determinant to find the truncations of the algebra. It is enough to identify these primaries and find zeros of their two point functions. We choose the normalization such that $W_{2j} = V_{2j} + \ldots$. With this choice, the two-point function of these fields is
\begin{align}
\label{prim2pt}
\nonumber
\langle W_2 W_2 \rangle & = -\frac{1}{2} n \left(2 \alpha_0^2+4 \alpha_0^2 n^2-6 \alpha_0^2 n-1\right) \\
\nonumber
\langle W_4 W_4 \rangle & = -\frac{1}{2 \left(20 \alpha_0^2 n^3-30 \alpha_0^2 n^2+10 \alpha_0^2 n-5 n-22\right)} \times n (2 n-1) \left(\alpha_0^2 n^2+\alpha_0^2 n-1\right) \times \\
\nonumber
& \times \left(12 \alpha_0^2+4 \alpha_0^2 n^2-14 \alpha_0^2 n-1\right) \left(2 \alpha_0^2+4 \alpha_0^2 n^2-6 \alpha_0^2 n-1\right) \left(-2 \alpha_0^2+4 \alpha_0^2 n^2-2 \alpha_0^2 n-9\right) \\
\nonumber
\langle W_6 W_6 \rangle & \sim (n-1) n (2 n-1) \left(\alpha_0^2 n^2+\alpha_0^2 n-1\right) \left(\alpha_0^2 n^2+2 \alpha_0^2 n-4\right) \left(30 \alpha_0^2+4 \alpha_0^2 n^2-22 \alpha_0^2 n-1\right) \times \\
\nonumber
& \times \left(12 \alpha_0^2+4 \alpha_0^2 n^2-14 \alpha_0^2 n-1\right) \left(2 \alpha_0^2+4 \alpha_0^2 n^2-6 \alpha_0^2 n-1\right) \left(4 \alpha_0^2 n^2-2 \alpha_0^2 n-1\right) \times \\
& \times \left(-2 \alpha_0^2+4 \alpha_0^2 n^2-2 \alpha_0^2 n-9\right) \left(-6 \alpha_0^2+4 \alpha_0^2 n^2+2 \alpha_0^2 n-25\right) \\
\nonumber
\langle W_8 W_8 \rangle & \sim (n-1) n (n+1) (2 n-3) (2 n-1) \left(n^2 \alpha_0^2+n \alpha_0^2-1\right) \left(n^2 \alpha_0^2+2 n \alpha_0^2-4\right) \times \\
\nonumber
& \times \left(n^2 \alpha_0^2+3 n \alpha_0^2-9\right) \left(4 n^2 \alpha_0^2-\alpha_0^2-4\right) \left(4 n^2 \alpha_0^2-30 n \alpha_0^2+56 \alpha_0^2-1\right) \times \\
\nonumber
& \times \left(4 n^2 \alpha_0^2-22 n \alpha_0^2+30 \alpha_0^2-1\right) \left(4 n^2 \alpha_0^2-14 n \alpha_0^2+12 \alpha_0^2-1\right) \times \\
\nonumber
& \times \left(4 n^2 \alpha_0^2-10 n \alpha_0^2+4 \alpha_0^2-9\right) \left(4 n^2 \alpha_0^2-6 n \alpha_0^2+2 \alpha_0^2-1\right) \left(4 n^2 \alpha_0^2-2 n \alpha_0^2-1\right) \times \\
\nonumber
& \times \left(4 n^2 \alpha_0^2-2 n \alpha_0^2-2 \alpha_0^2-9\right) \left(4 n^2 \alpha_0^2+2 n \alpha_0^2-6 \alpha_0^2-25\right) \left(4 n^2 \alpha_0^2+6 n \alpha_0^2-10 \alpha_0^2-49\right)
\end{align}
where $\sim$ means that we didn't write the denominator (because we are mainly interested in zeros of these two-point functions). One can actually find higher order two-point function by the following trick: we have
\begin{equation}
C_{46}^8 C_{88}^0 = C_{48}^6 C_{66}^0
\end{equation}
if the field $W_8$ is chosen to be orthogonal to $W_{[44]}$ which is equivalent to condition
\begin{equation}
C_{48}^4 = 0.
\end{equation}
Similarly at the next level
\begin{equation}
C_{48}^{10} C_{10,10}^0 = C_{4,10}^8 C_{88}^0
\end{equation}
if we choose $W_{10}$ to be orthogonal to $W_{[46]}$ and $W_{[44]^{(2)}}$ which means
\begin{equation}
C_{4,10}^4 = 0 \quad\quad \text{and} \quad \quad C_{4,10}^6 = 0.
\end{equation}
In this way, we were able to find the zeros of Kac determinant up to level $12$ which would otherwise require to knowing $24$th order pole of $W_{12}$ with itself. The fact that the numerator of $C_{10,10}^0$ and $C_{12,12}^0$ obtained in this way factorizes into factors of the form of (\ref{prim2pt}) is a nice check of consistency of this procedure.

\section{Truncations}
We will now collect all the results about truncations using various truncations discussed so far. All the truncation curves will have formally the same form as in $\mathcal{W}_\infty$,
\begin{equation}
\label{trunccurve}
\frac{N_1}{\mu_1} + \frac{N_2}{\mu_2} + \frac{N_3}{\mu_3} = 1
\end{equation}
with non-negative integers $N_1, N_2$ and $N_3$. Due to redundancy in parametrization (\ref{muconstraint}) shifting all $N_j$ by a constant does not change the truncation curve, but we can use these triples of integers differing by a constant to describe different truncations of the algebra with the same truncation curve.

\subsection{Truncations from Gaiotto-Rap\v{c}\'{a}k algebras}
The expressions for $\eta$ for $Y$-algebras discussed in section \ref{secgr} can be immediately translated into truncation curves, i.e. curves in $\mu$-parameter space where the universal even spin $\mathcal{W}_\infty$ truncates to a smaller algebra. These curves have the form
\begin{align}
\nonumber
Y^-: & \quad \frac{2N_1+1}{\mu_1} + \frac{2N_2+1}{\mu_2} + \frac{2N_3}{\mu_3} = 1 \\
\nonumber
\tilde{Y}^-: & \quad \frac{2N_1+1}{\mu_1} + \frac{2N_2}{\mu_2} + \frac{2N_3+1}{\mu_3} = 1 \\
Y^+: & \quad \frac{2N_1}{\mu_1} + \frac{2N_2+1}{\mu_2} + \frac{2N_3+1}{\mu_3} = 1 \\
\nonumber
\tilde{Y}^+: & \quad \frac{2N_1}{\mu_1} + \frac{2N_2}{\mu_2} + \frac{2N_3}{\mu_3} = 1
\end{align}
These have a very simple form: if the gauge group associated to face $\mu_j$ is $Sp(2N)$, the coefficient of $\mu_j^{-1}$ is $2N$, while if the gauge group is $SO(N)$, the coefficient is $N-1$ (both for even and odd $N$). Whenever the parameters $(\mu_1,\mu_2,\mu_3)$ of the even spin $\mathcal{W}_\infty$ satisfy one of these equations for non-negative integer values of $(N_1,N_2,N_3)$, the algebra develops an ideal so can be truncated to a smaller subalgebra. Note that in general for a fixed truncation curve there might be various choices of this ideal corresponding to the fact that the map from triples $(N_1,N_2,N_3)$ to truncation curves is not one-to-one (in particular an overall shift of all three ranks by a constant leads to the same truncation curve). In the case of $\mathcal{W}_\infty$ there was always a maximal ideal corresponding to a truncation where (at least) one of the integers $N_j$ was vanishing. One way to understand what is happening is to analyze the characters of $Y$-algebras and study the level at which the first singular vector appears.

\subsection{Truncations from cosets and DS reductions}
We can similarly translate the value of $\eta$ for cosets and Drinfe\v{l}d-Sokolov reductions to truncation curves. The orthogonal cosets (\ref{socoset}) have simple truncation curves
\begin{equation}
\frac{n-1}{\mu_1} = 1
\end{equation}
while the symplectic ones
\begin{equation}
\frac{2n+1}{\mu_2} + \frac{2n+1}{\mu_3} = 1.
\end{equation}
The Drinfe\v{l}d-Sokolov reductions lead to curves
\begin{align}
\nonumber
B_n: & \quad \frac{1}{\mu_1} + \frac{2n_B+1}{\mu_3} = 1 \\
C_n: & \quad \frac{1}{\mu_2} + \frac{2n_C+1}{\mu_3} = 1 \\
\nonumber
D_n: & \quad \frac{2n_D-1}{\mu_3} = 1.
\end{align}

\paragraph{Unitary minimal models}
These of course don't exhaust all possible truncations that we may get by studying cosets and Drinfe\v{l}d-Sokolov reductions. For example cosets (\ref{socoset}) can be studied at fixed non-negative integer value of $k$ and generic $n$. The central charges of these models are $c=0$ ($k=0$, only the vacuum state), $c=1$ ($k=1$), $c=\frac{4n-1}{2n+1}$ ($k=2$), \ldots. These are the unitary minimal models of the corresponding truncated algebras. For each $k$, the parameters of the associated even spin $\mathcal{W}_\infty$ lie on a curve
\begin{equation}
\frac{k}{\mu_2} + \frac{k-1}{\mu_3} = 1
\end{equation}
so we can think of this curve as cutting out the unitary minimal models in the parameter space. This is again very similar to the situation in $\mathcal{W}_{1+\infty}$ and in fact even the form of these curves is the same.

\paragraph{Non-unitary minimal models}
The class of all minimal models is larger than one with unitary minimal models. Consider following \cite{Lukyanov:1989gg} the minimal models of $\mathcal{W}$-algebra associated to $D_n$ via Drinfe\v{l}d-Sokolov reduction parametrized by coprime integers $(p^\prime,p)$ such that the central charge is
\begin{equation}
c = n \left[ 1 - 2(n-1)(2n-1) \frac{(p^\prime-p)^2}{p^\prime p} \right].
\end{equation}
Choosing $p^\prime = p+1$ and $p = 2n-2+k$ we get for $k=0,1,\ldots$ the sequence of unitary minimal models discussed in the previous paragraph. For $|p^\prime-p| \neq 1$ we still get minimal models but no longer unitary. The level of Drinfe\v{l}d-Sokolov reduction can be chosen either
\begin{equation}
k_{D} = \frac{p^\prime-2(n-1)p}{p} \quad\quad\text{or}\quad\quad \frac{p-2(n-1)p^\prime}{p^\prime}.
\end{equation}
Choosing the first one, we can identify the parameters of even spin $\mathcal{W}_\infty$ as
\begin{align}
\nonumber
\mu_1 & = \frac{p}{(2n-1)(p^\prime-p)} \\
\mu_2 & = -\frac{p^\prime}{(2n-1)(p^\prime-p)} \\
\nonumber
\mu_3 & = \frac{1}{2n-1}.
\end{align}
These lie on truncation curve
\begin{equation}
\frac{p^\prime-2n+1}{\mu_1} + \frac{p-2n+1}{\mu_2} = 1.
\end{equation}
Choosing $p^\prime-p=1$ we reduce to the truncation curve of minimal models discussed in the previous paragraph (although in another triality frame).

\subsection{Truncations from explicit bootstrap}
Let us summarize truncation curves that we see from the explicit calculation of operator product expansions. This is easier to see in the quadratic basis because we have a natural normalization of fields such that the OPEs have only polynomial coefficients in this basis.

\paragraph{Truncation to vacuum}
Just like the $c=0$ truncation of Virasoro algebra where the vacuum representation is one-dimensional, in even spin $\mathcal{W}_\infty$ for
\begin{equation}
\frac{1}{\mu_1} + \frac{1}{\mu_2} = 1
\end{equation}
(and permutations of $\mu$) the dimension two field $V_2$ is singular so the theory reduces to a single state. This happens for example in the zeroth unitary minimal model $k=0$ where there is just the vacuum state and $c=0$.

\paragraph{Truncation to $W[2]$ (Virasoro)}
The Virasoro algebra generated by $T = -V_2$ is always a subalgebra of even spin $\mathcal{W}_\infty$. If we are interested in quotient algebras and the corresponding ideals, the dimension $4$ field is singular only if equation of the form
\begin{equation}
\frac{1}{\mu_1} + \frac{3}{\mu_2} = 1
\end{equation}
is satisfied. In this case the singular vector is at level $4$. These truncations of even spin $\mathcal{W}_\infty$ admit free field representation in terms of only one free boson.

\paragraph{Truncation to $W[2,4]$}
Working in primary basis, truncation to $\mathcal{W}$-algebra with additional spin $4$ field is a little bit irregular because our parameter $x$ is not defined. The only condition coming from associativity of the algebra is
\begin{equation}
\frac{C_{44}^0}{(C_{44}^4)^2} = \frac{c(2c-1)(5c+22)(7c+68)}{216(c+24)(c^2-172c+196)}
\end{equation}
Translated to truncation curves, we find three curves of the form
\begin{equation}
\frac{1}{\mu_1} = 1
\end{equation}
(which corresponds to first unitary minimal models) and we also have an orbit of six curves of the form (these are associated to $WB_2$ or $WC_2$ truncations)
\begin{equation}
\frac{1}{\mu_1} + \frac{5}{\mu_2} = 1.
\end{equation}
All these algebras have level $6$ singular vector.

\paragraph{Truncation to $W[2,4,6]$}
The bootstrap for algebras of type $W[2,4,6]$ is consistent if $x$ takes one of the values
\begin{equation}
\frac{5(2c-1)(3c+20)(7c+68)}{6(c+24)(10c^2+47c-82)}, \quad\quad \frac{5(c+50)(2c-1)(7c+68)}{3(c+24)(5c^2+309c-14)}
\end{equation}
as well as one of two roots of the quadratic equation
\begin{equation}
a_2 x^2 + a_1 x + a_0 = 0
\end{equation}
with
\begin{align}
\nonumber
a_2 & = 18(c+24)^2(85c^4+5275c^3+101736c^2+1806268c-2633664) \\
a_1 & = 3(c+24)(7c+68)(65c^4+2409c^3-161760c^2-11131676c+17536992) \\
\nonumber
a_0 & = 98(c+50)(2c-1)(7c+68)^2(13c+1320)
\end{align}
The first solution for $x$ corresponds to truncations
\begin{equation}
\frac{2}{\mu_1} = 1
\end{equation}
(and triality images of this), the second solution to
\begin{equation}
\frac{3}{\mu_1} + \frac{3}{\mu_2} = 1
\end{equation}
and the pair of algebraic solutions satisfying the quadratic equation for $x$ correspond to six truncation curves of the form
\begin{equation}
\frac{1}{\mu_1} + \frac{7}{\mu_2} = 1
\end{equation}
(these are the $WB_3$ or $WC_3$ truncations). There are also some spurious co-dimension two specializations of parameters where the algebra truncates (for example $(c=-\frac{22}{5},x=\frac{31}{28})$ or $(c=-\frac{68}{7},x=0)$ but we are interested in co-dimension $1$ specializations so we don't discuss these.

\paragraph{Truncation to $W[2,4,6,8]$}
Here the truncation curves are of the form
\begin{equation}
\frac{3}{\mu_1} = 1
\end{equation}
and
\begin{equation}
\frac{1}{\mu_1} + \frac{9}{\mu_2} = 1.
\end{equation}

\paragraph{Truncation to $W[2,4,6,8,10]$}
At level $12$ there are four different types of truncations,
\begin{equation}
\frac{4}{\mu_1} = 1, \quad\quad \frac{2}{\mu_1}+\frac{1}{\mu_2} = 1, \quad\quad \frac{5}{\mu_1} + \frac{3}{\mu_2} = 1, \quad\quad \frac{11}{\mu_1} + \frac{1}{\mu_2} = 1.
\end{equation}

\subsection{Summary up to level $12$ and conjecture}
Let's summarize the truncations discussed in this section. The following table lists all the truncations with singular vector up to level $12$:
\begin{center}
\begin{tabular}{|c|c|c|}
\hline
$(N_1,N_2,N_3)$ & level of singular vector & construction of truncation \\
\hline
$(1,1,0)$ & $2$ & vacuum \\
$(3,1,0)$ & $4$ & Virasoro \\
$(1,0,0)$ & $6$ & first unitary minimal models \\
$(5,1,0)$ & $6$ & $WB_2 \simeq WC_2$ \\
$(2,0,0)$ & $8$ & $\mathfrak{so}(3)$ coset \\
$(7,1,0)$ & $8$ & $WB_3 \simeq WC_3$ \\
$(3,3,0)$ & $8$ & $\mathfrak{sp}(2)$ coset \\
$(3,0,0)$ & $10$ & $WD_2$, $\mathfrak{so}(4)$ coset \\
$(9,1,0)$ & $10$ & $WB_4 \simeq WC_4$ \\
$(4,0,0)$ & $12$ & $\mathfrak{so}(5)$ coset \\
$(2,1,0)$ & $12$ & second unitary minimal models \\
$(5,3,0)$ & $12$ & \\
$(11,1,0)$ & $12$ & $WB_5 \simeq WC_5$ \\
\hline
\end{tabular}
\end{center}
To write a general conjecture for a level of a given truncation we need to distinguish three cases depending on the even/odd parity of the parameters $N_j$ in (\ref{trunccurve}):
\begin{enumerate}
\item For the truncation curves of the form
\begin{equation}
\frac{2N_1+1}{\mu_1} + \frac{2N_2+1}{\mu_2} = 1
\end{equation}
the truncation has first singular vector at level
\begin{equation}
\frac{1}{2} (2N_1+2) \times (2N_2+2) \times 1.
\end{equation}
In Gaiotto-Rap\v{c}\'{a}k picture this corresponds to one of algebras $Y^-_{N_1,N_2,0}$, $Y^-_{N_2,N_1,0}$, $\tilde{Y}^-_{N_1,0,N_2}$, $\tilde{Y}^-_{N_2,0,N_1}$, $Y^+_{0,N_1,N_2}$ or $Y^+_{0,N_2,N_1}$. In each of these cases we have gauge groups $Sp(2N_1)$ or $SO(2N_1+1)$ and $Sp(2N_2)$ or $SO(N_2+1)$. The third gauge group is formally $SO(0)$.
\item Second type of truncation curves are those of the form
\begin{equation}
\frac{2N_1+1}{\mu_1} + \frac{2N_2}{\mu_2} = 1.
\end{equation}
These truncations have their first singular vector at level
\begin{equation}
\frac{1}{2} (2N_1+3) \times (2N_2+2) \times 2
\end{equation}
and the Gaiotto-Rap\v{c}\'{a}k algebras are now $Y^-_{0,N_2,N_1+1}$, $Y^-_{N_2,0,N_1+1}$, $\tilde{Y}^-_{0,N_1+1,N_2}$, $\tilde{Y}^-_{N_2,N_1+1,0}$, $Y^+_{N_1+1,0,N_2}$ or $Y^+_{N_1+1,N_2,0}$. The associated gauge groups are $SO(2N_1+2)$, either $Sp(2N_2)$ or $SO(2N_2+1)$ and formally $Sp(0)$.
\item The last type of truncation curves are those of the form
\begin{equation}
\frac{2N_1}{\mu_1} + \frac{2N_2}{\mu_2} = 0.
\end{equation}
The $Y$-algebras are of the form $\tilde{Y}^+_{N_1,N_2,0}$ and permutations and the gauge groups are either $Sp(2N_1)$ or $SO(2N_1+1)$ and either $Sp(2N_2)$ or $SO(2N_2+1)$. The third gauge group is formally $Sp(0)$. The level of such truncations is
\begin{equation}
\frac{1}{2} (2N_1+2) \times (2N_2+2) \times 2
\end{equation}
\end{enumerate}
The level of the truncation is now given uniformly as
\begin{equation}
\label{trunclevel}
\frac{1}{2} \rho(G_1) \times \rho(G_2) \times \rho(G_3)
\end{equation}
where $\rho(G)$ is an independent factor associated to each gauge group,
\begin{equation}
\rho(G) = \begin{cases} 2n+2, \quad Sp(2n) \\ 2n+2, \quad SO(2n+1) \\ 2n+1, \quad SO(2n) \end{cases}
\end{equation}
or in other words twice the (Dynkin) rank plus the lacity ($1$ for simply laced $D_n$ and $2$ for doubly laced algebras $B_n$ and $C_n$).

We explicitly verified these truncation curves only by studying the first appearance of the singular vector in the universal even spin algebra. In this way we can only detect the truncations where one of the $N_j$ parameters vanishes. This corresponds to simple quotients of the algebra. The class of $Y$-algebras introduced in \cite{Gaiotto:2017euk} however includes also algebras which are not simple. These are still interesting for example when one considers the gluing \cite{Prochazka:2017qum} because in general a simple algebra can obtained by gluing of non-simple subalgebras. In the unitary case the free field representations of these non-simple quotients were found in \cite{Prochazka:2018tlo}. Since in the unitary case which is better understood and also in all examples discussed here the level of the first singular vector follows a simple uniform factorized formula (\ref{trunclevel}) where the individual gauge groups don't interact and which makes good sense even if all parameters parametrizing the truncation curve are non-zero, we conjecture that this correctly describes the truncation of $Y$-algebras in the non-simple situation as well.

\paragraph{Comparison of truncation curves of even spin $\mathcal{W}_\infty$ and $\mathcal{W}_\infty$}
In general each truncation curve
\begin{equation}
\frac{N_1}{\mu_1} + \frac{N_2}{\mu_2} + \frac{N_3}{\mu_3} = 1
\end{equation}
in even spin $\mathcal{W}_\infty$ lies on a curve
\begin{equation}
\frac{N_1}{2\lambda_1} + \frac{N_2}{2\lambda_2} + \frac{N_3+1}{2\lambda_3} = 1.
\end{equation}
in the parameter space of $\mathcal{W}_\infty$. Due to factor of $2$ in the denominator there are curves in the parameter space of $\mathcal{W}_\infty$ where the full algebra does not truncate but the even spin subalgebra can still truncate. Truncations to algebras $WB_n, WC_n$ and $WD_n$ are examples of truncations which lie on truncation curves in $\mathcal{W}_{1+\infty}$, actually they lie on curves corresponding to $W_n$ algebras (this is also true for exceptional algebras where the embeddings in $\mathcal{W}_{1+\infty}$ are known).

\section{Gluing}
In this last section we illustrate how the gluing procedure discussed in \cite{Prochazka:2017qum} applies to orthogonal affine Lie algebras. Let us first review the case of unitary affine Lie algebras. The gluing diagram of $\mathfrak{u}(N)_k$ is based on the decomposition
\begin{equation}
\mathfrak{u}(N)_k \supset \frac{\mathfrak{u}(N)_k}{\mathfrak{u}(N-1)_k} \times \frac{\mathfrak{u}(N-1)_k}{\mathfrak{u}(N-2)_k} \times \cdots \times \frac{\mathfrak{u}(2)_k}{\mathfrak{u}(1)} \times \mathfrak{u}(1).
\end{equation}
Each of the factors on the right hand side is a truncation of $\mathcal{W}_{1+\infty}$.
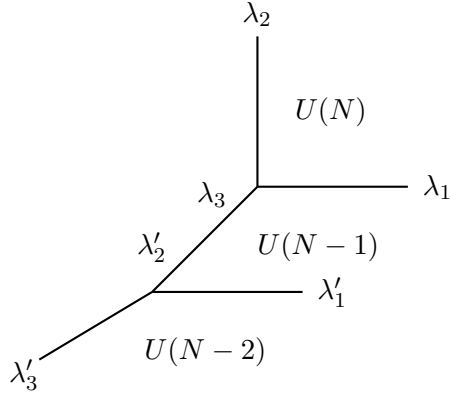
\begin{figure}
\begin{center}
\begin{tikzpicture}
\draw[thick] (0,0) -- (2,0);
\draw[thick] (0,0) -- (0,2);
\draw[thick] (0,0) -- (-1.4,-1.4);
\draw[thick] (-1.4,-1.4) -- (0.6,-1.4);
\draw[thick] (-1.4,-1.4) -- (-2.9,-2.3);
\node[text width=3cm] at (2,1) {$U(N)$};
\node[text width=3cm] at (1.5,-0.8) {$U(N-1)$};
\node[text width=3cm] at (0,-2.2) {$U(N-2)$};
\node[text width=3cm] at (1.3,2.3) {$\lambda_2$};
\node[text width=3cm] at (3.7,0) {$\lambda_1$};
\node[text width=3cm] at (2.3,-1.4) {$\lambda_1^\prime$};
\node[text width=3cm] at (-1.8,-2.5) {$\lambda_3^\prime$};
\node[text width=3cm] at (0.7,-0.1) {$\lambda_3$};
\node[text width=3cm] at (-0.1,-0.8) {$\lambda_2^\prime$};
\end{tikzpicture}
\end{center}
\caption{Part of gluing diagram for $\mathfrak{u}(N)_k$.}
\label{figunitary}
\end{figure}
Identifying parameters as in figure \ref{figunitary} we can calculate the $\lambda$-parameters of the corresponding $\mathcal{W}_{1+\infty}$ algebras sitting at the vertices \cite{Prochazka:2017qum}. We have
\begin{align}
\lambda_3 & = \frac{(N-1)\epsilon_2+N\epsilon_3}{\epsilon_3} = \frac{N(\psi-1)-(N-1)\psi}{\psi-1}, \\
\lambda_2^\prime & = \frac{(N-2)\epsilon_2^\prime+(N-1)\epsilon_3^\prime}{\epsilon_2^\prime} = -\frac{(N-1)(\psi^\prime-1)-(N-2)\psi^\prime}{\psi^\prime}
\end{align}
From the $(p,q)$ charges of the five-branes we see that $\epsilon_1^\prime = \epsilon_1$ and $\epsilon_2^\prime = \epsilon_1+\epsilon_2$ so
\begin{equation}
\label{glueor}
\psi^\prime \equiv -\frac{\epsilon_2^\prime}{\epsilon_1^\prime} = \psi-1
\end{equation}
which guarantees that the dimension of the fundamental gluing fields is
\begin{equation}
h = h_3 + h_2^\prime = \frac{1+\lambda_3}{2} + \frac{1+\lambda_2^\prime}{2} = 1
\end{equation}
This is exactly what we need in order to find dimension $1$ fields charged under Cartan $\mathfrak{u}(1)$ currents coming from the vertices. These correspond in the language of affine Lie algebras to currents associated to positive and negative simple roots. The other generators associated to roots that are not simple corresponding to line operators stretched between vertices which are not neighbouring.

Let's now consider the orthogonal Lie algebra $\mathfrak{so}(2n+1)_k$. We have a similar decomposition
\begin{equation}
\mathfrak{so}(2N+1)_k \supset \frac{\mathfrak{so}(2N+1)_k}{\mathfrak{so}(2N)_k} \times \frac{\mathfrak{so}(2N)_k}{\mathfrak{so}(2N-1)_k} \times \cdots \times \frac{\mathfrak{so}(3)_k}{\mathfrak{so}(2)} \times \mathfrak{so}(2).
\end{equation}
The first coset on the right hand side has parameters compatible with $\tilde{Y}^-_{0,N,N}$ with the truncation curve
\begin{equation}
\frac{2N}{\mu_2} + \frac{2N+1}{\mu_3} = 1
\end{equation}
while the second term can be identified with $Y^-_{0,N-1,N}$ with truncation curve
\begin{equation}
\frac{2N-2}{\mu_2^\prime} + \frac{2N-1}{\mu_3^\prime} = 1.
\end{equation}
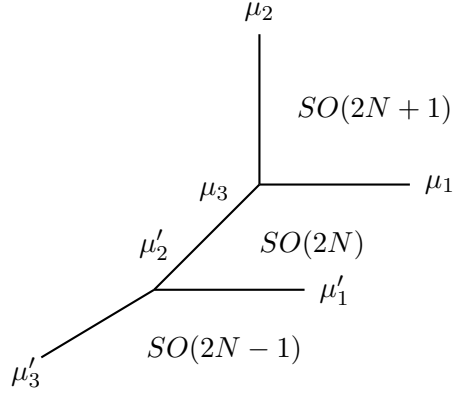
\begin{figure}
\begin{center}
\begin{tikzpicture}
\draw[thick] (0,0) -- (2,0);
\draw[thick] (0,0) -- (0,2);
\draw[thick] (0,0) -- (-1.4,-1.4);
\draw[thick] (-1.4,-1.4) -- (0.6,-1.4);
\draw[thick] (-1.4,-1.4) -- (-2.9,-2.3);
\node[text width=3cm] at (2,1) {$SO(2N+1)$};
\node[text width=3cm] at (1.5,-0.8) {$SO(2N)$};
\node[text width=3cm] at (0,-2.2) {$SO(2N-1)$};
\node[text width=3cm] at (1.3,2.3) {$\mu_2$};
\node[text width=3cm] at (3.7,0) {$\mu_1$};
\node[text width=3cm] at (2.3,-1.4) {$\mu_1^\prime$};
\node[text width=3cm] at (-1.8,-2.5) {$\mu_3^\prime$};
\node[text width=3cm] at (0.7,-0.1) {$\mu_3$};
\node[text width=3cm] at (-0.1,-0.8) {$\mu_2^\prime$};
\end{tikzpicture}
\end{center}
\caption{Part of gluing diagram for $\mathfrak{so}(2N+1)_k$.}
\label{figorthogonal}
\end{figure}
The first part of the gluing diagram looks like figure \ref{figorthogonal}. Let us verify that the gluing fields have compatible dimensions. The upper vertex has parameter
\begin{equation}
\mu_3 = \frac{\psi-2N}{\psi_1}
\end{equation}
while the corresponding parameter of the lower vertex is
\begin{equation}
\mu_2^\prime = -\frac{1-2N+\psi^\prime}{\psi^\prime}.
\end{equation}
The relative orientation of the two vertices is just like in the unitary case so we still have (\ref{glueor}). Now we can calculate the conformal dimension of gluing fields and find
\begin{equation}
h_3 + h_2^\prime = 1
\end{equation}
which is exactly what we want in order to find dimension $1$ currents.

Note that they way the currents appear is slightly different than in the unitary situation. In the unitary case each vertex represented a truncation of $\mathcal{W}_{1+\infty}$ algebra which by definition carried an affine $\mathfrak{u}(1)$ current. There are as many of these as is the rank of the algebra and all these currents give the Cartan subalgebra of $\mathfrak{u}(N)_k$. As already discussed the elementary gluing fields give rise to simple positive and negative roots. In the orthogonal case the truncations of even spin $\mathcal{W}_\infty$ algebras at vertices do not have any spin $1$ fields so we don't find any Cartan fields in this way. On the other hand, we have an alternating sequence of $Y^-$ and $\tilde{Y}^-$ algebras and associated to each neighbouring pair of these there is an elementary dimension $1$ gluing field (which now do not appear in complex conjugate pairs because the minimal representations of even spin $\mathcal{W}_\infty$ are real). For example for rank $2$ algebra the first Cartan generator can be chosen to correspond to line operator stretched from $\tilde{Y}^-_{022}$ to $Y^-_{012}$ and the second generator to line operator between $\tilde{Y}^-_{011}$ to $Y^-_{001}$.

\section{Discussion}
The understanding of the universal orthosymplectic $\mathcal{W}_\infty$ algebra is still much more limited than that of $\mathcal{W}_{1+\infty}$. In particular
\begin{enumerate}
\item All the truncations that we found are associated to truncation curves of the form (\ref{trunccurve}) and also each of these can be associated to a certain $Y$-algebra. We conjecture that this happens at level (\ref{trunclevel}), but our calculations give no proof that this is what actually happens. From the from explicit coset and Drinfe\v{l}d-Sokolov reduction description of $Y$-algebras it should be possible to verify this.
\item The free field representations of truncations of $\mathcal{W}_{1+\infty}$ are reasonably well understood \cite{Prochazka:2018tlo}. Since the even spin algebra is a subalgebra of $\mathcal{W}_{1+\infty}$, we can find many free field representations of truncations of even spin algebra from the free field representations of $\mathcal{W}_{1+\infty}$. But one should understand if there are any other representations and how are these related to truncations of the algebra, i.e. if we have a correspondence between free field representations and co-dimension $1$ truncations like in the case of $\mathcal{W}_{1+\infty}$ \cite{Prochazka:2018tlo}.
\item The combinatorial box counting interpretation of characters of even spin $\mathcal{W}_\infty$ is not known. One cannot simply restrict to subset of box configurations in $\mathcal{W}_{1+\infty}$ because the canonical Virasoro generators don't agree and the higher spin generators of $\mathcal{W}_{1+\infty}$ seem not to preserve the even spin subalgebra.
\item No analogue of Tsymbaliuk presentation of $\mathcal{W}_{1+\infty}$ as affine Yangian is known. One could try to repeat the steps of \cite{Prochazka:2019dvu} to find the ladder operators in Yangian but first the folding of $GL(N)$ Miura operator should be understood. It is very reminiscent to spin chains with boundary where the boundary reflection operator is $\partial$ or $\partial^{-1}$. This surely deserves a deeper study. Once this is understood one can try to apply the techniques of quantum inverse scattering method or algebraic Bethe ansatz to construct Yangian operators for even spin $\mathcal{W}_{1+\infty}$.
\item Although the algebra admits a quadratic basis, the derivatives of fields don't seem to follow the same simple pattern as in the case of $\mathcal{W}_{1+\infty}$. If one understands this, one might hope to be able to write a closed-form formulas for OPEs and commutators in even spin $\mathcal{W}_\infty$ just like those in \cite{Prochazka:2014gqa,Eberhardt:2019xmf}.
\item In $\mathcal{W}_{1+\infty}$ and its matrix extension the fusion and its associated coproduct were extremely efficient tools for construction of free field representations or representations in terms of affine Lie algebras. Also the space of co-dimension $1$ truncations can be seen as a cone generated by elementary Miura transformations \cite{Eberhardt:2019xmf,Rapcak:2019wzw}. The unitary Miura operator immediately allows us to extract the coproduct. On the other hand, because of the folding of the Miura operator in the orthosymplectic case, it is not obvious if the orthosymplectic version of $\mathcal{W}_\infty$ admits this coproduct structure.
\end{enumerate}

\section*{Acknowledgements}

I would like to thank to Lorenz Eberhardt, Andrew Linshaw and Miroslav Rap\v{c}\'{a}k for useful discussions. This research was supported by the DFG Transregional Collaborative Research Centre TRR 33 and the DFG cluster of excellence Origin and Structure of the Universe.

\appendix

\section{Structure constants in primary basis}
\label{primaryeqns}
Here is the list of the structure constants in the primary basis for sum of spins up to 12:
\begin{align*}
C_{44}^0 & = \frac{c(5c+22)}{72(c+24)} \left(C_{44}^4\right)^2 -\frac{7c(c-1)(5c+22)}{72(2c-1)(7c+68)} C_{44}^4 C_{46}^6 + \frac{c(c-1)(c+24)(5c+22)}{12(2c-1)(7c+68)^2} \left(C_{46}^6\right)^2 \\
C_{46}^4 & = \frac{4(5c+22)}{9(c+24)} \frac{\left(C_{44}^4\right)^2}{C_{44}^6} - \frac{96(c^2-172c+196)}{c(2c-1)(7c+68)} \frac{C_{44}^0}{C_{44}^6} \\
C_{48}^4 & = -\frac{32(c-151)(c-1)(c+24)(5c+22)(c^2-172c+196)}{(2c-1)^2(7c+68)^3(13c+516)} \frac{\left(C_{46}^6\right)^3}{C_{44}^6 C_{46}^8} \\
& - \frac{56(c-1)(5c+22)(c^2-172c+196)(20c^3+24807c^2+765640c-185172)}{3(c+31)(2c-1)^2(7c+68)^2(13c+516)(55c-6)} \frac{C_{44}^4 \left(C_{46}^6\right)^2}{C_{44}^6 C_{46}^8} \\
& + \frac{4(c-1)(5c+22)(5605c^4-408494c^3-70820464c^2-1703657536c+1312613664)}{9(c+24)(c+31)(2c-1)(7c+68)(13c+516)(55c-6)} \frac{\left(C_{44}^4\right)^2 C_{46}^6}{C_{44}^6 C_{46}^8} \\
& -\frac{5(c-1)(c+50)(5c+22)(715c^4+90933c^3+2851076c^2+21154896c+6967008)}{12(c+24)(c+31)(3c+46)(5c+3)(13c+516)(55c-6)} \frac{C_{46}^{[44]} \left(C_{44}^4\right)^2}{C_{46}^8} \\
& + \frac{7(c-1)(5c+22)(65c^4+8637c^3+364470c^2+2897944c+36384)}{12(2c-1)(3c+46)(5c+3)(7c+68)(13c+516)} \frac{C_{46}^{[44]} C_{44}^4 C_{46}^6}{C_{46}^8} \\
& - \frac{(c-1)(c+24)(5c+22)(65c^4+8637c^3+364470c^2+2897944c+36384)}{2(2c-1)(3c+46)(5c+3)(7c+68)^2(13c+516)} \frac{C_{46}^{[44]} \left(C_{46}^6\right)^2}{C_{46}^8} \\
& +\frac{140(c-1)(c+50)(5c+22)(11c+656)}{27(c+24)^2(c+31)(55c-6)} \frac{\left(C_{44}^4\right)^3}{C_{44}^6 C_{46}^8} \\
C_{48}^6 & = \frac{8(25c^3+615c^2-88272c+102332)}{3(c+24)(c+31)(55c-6)} \frac{C_{44}^4 C_{46}^6}{C_{46}^8} -\frac{4(c-151)}{(13c+516)} \frac{C_{46}^{[44]} C_{44}^6 C_{46}^6}{C_{46}^8} \\
& + \frac{16(425c^4+15145c^3+233766c^2+6507708c-7565544)}{(c+31)(7c+68)(13c+516)(55c-6)} \frac{\left(C_{46}^6\right)^2}{C_{46}^8}\\
& +\frac{7840(c+50)(2c-1)(7c+68)}{9(c+24)^2(c+31)(55c-6)} \frac{\left(C_{44}^4\right)^2}{C_{46}^8} -\frac{35(c+50)(2c-1)(7c+68)}{(3c+24)(c+31)(55c-6)} \frac{C_{46}^{[44]} C_{44}^4 C_{44}^6}{C_{46}^8} \\
C_{48}^8 & = \frac{8(33c^2+1087c+11760)}{(7c+68)(13c+516)} C_{46}^6 - \frac{(31c-192)}{2(13c+516)} C_{44}^6 C_{46}^{[44]} -2C_{44}^4 \\
C_{48}^{[44]} & = -\frac{896(3c+46)(5c+3)(c^2-172c+196)}{(c+31)(2c-1)(7c+68)^2(55c-6)} \frac{\left(C_{46}^6\right)^2}{C_{44}^6 C_{46}^8} + \frac{8(33c^2+1087c+11760)}{(7c+68)(13c+516)} \frac{C_{46}^6 C_{46}^{[44]}}{C_{46}^8} \\
& +\frac{3136(3c+46)(5c+3)(c^2-172c+196)}{3(c+24)(c+31)(2c-1)(7c+68)(55c-6)} \frac{C_{44}^4 C_{46}^6}{C_{44}^6 C_{46}^8} -\frac{(31c-192)}{2(13c+516)} \frac{C_{44}^6 \left(C_{46}^{[44]}\right)^2}{C_{46}^8} \\
& -\frac{4(165c^3+10763c^2+140036c+38568)}{3(c+24)(c+31)(55c-6)} \frac{C_{44}^4 C_{46}^{[44]}}{C_{46}^8} +\frac{448(3c+46)(5c+3)(11c+656)}{9(c+24)^2(c+31)(55c-6)} \frac{\left(C_{44}^4\right)^2}{C_{44}^6 C_{46}^8} \\
C_{48}^{[46]^{(1)}} & = -\frac{112}{3(c+24)} \frac{C_{44}^4}{C_{46}^8} + \frac{(13c+918)}{(13c+516)} \frac{C_{44}^6 C_{46}^{[44]}}{C_{46}^8} -\frac{16(113c-1338)}{(7c+68)(13c+516)} \frac{C_{46}^6}{C_{46}^8} \\
C_{66}^0 & = -\frac{7(c-1)c(5c+22)^2(8c^2+1161c-1244)}{162(c+24)(2c-1)^2(7c+68)^2} \frac{\left(C_{44}^4\right)^3 C_{46}^6}{\left(C_{44}^6\right)^2} \\
& +\frac{14(c-1)^2c(c+24)(5c+22)^2(c^2-172c+196)}{9(2c-1)^3(7c+68)^4} \frac{C_{44}^4 \left(C_{46}^6\right)^3}{\left(C_{44}^6\right)^2} \\
& -\frac{2(c-1)^2c(c+24)^2(5c+22)^2(c^2-172c+196)}{3(2c-1)^3(7c+68)^5} \frac{\left(C_{46}^6\right)^4}{\left(C_{44}^6\right)^2} \\
& -\frac{(c-1)c(5c+22)^2(17c^3-13105c^2+25330c-12092)}{54(2c-1)^3(7c+68)^3} \frac{\left(C_{44}^4\right)^2 \left(C_{46}^6\right)^2}{\left(C_{44}^6\right)^2} \\
& +\frac{(c-1)c(5c+22)^2(11c+656)}{162(c+24)^2(2c-1)(7c+68)} \frac{\left(C_{44}^4\right)^4}{\left(C_{44}^6\right)^2} \\
C_{66}^4 & = \frac{28(c-1)(5c+22)(c^2-172c+196)}{3(2c-1)^2(7c+68)^2} \frac{C_{44}^4 \left(C_{46}^6\right)^2}{\left(C_{44}^6\right)^2} \\
& -\frac{8(c-1)(c+24)(5c+22)(c^2-172c+196)}{(2c-1)^2(7c+68)^3} \frac{\left(C_{46}^6\right)^3}{\left(C_{44}^6\right)^2} \\
& +\frac{4(c-1)(5c+22)(11c+656)}{9(c+24)(2c-1)(7c+68)}\frac{\left(C_{44}^4\right)^2 C_{46}^6}{\left(C_{44}^6\right)^2} \\
C_{66}^6 & = -\frac{20(13c^4-1637c^3-113622c^2+32168c+859328)}{27(c+24)^2(2c-1)(7c+68)} \frac{\left(C_{44}^4\right)^2}{C_{44}^6} \\
& +\frac{10(28c^5-5425c^4-525974c^3+387728c^2+3726976c-3870208)}{9(c+24)(2c-1)^2(7c+68)^2} \frac{C_{44}^4 C_{46}^6}{C_{44}^6} \\
& +\frac{20(92c^5+2389c^4+39632c^3+4060c^2-212032c+193984)}{(2c-1)^2(7c+68)^3} \frac{\left(C_{46}^6\right)^2}{C_{44}^6} \\
C_{66}^8 & = \frac{4(4c+61)}{(7c+68)} \frac{C_{46}^6 C_{46}^8}{C_{44}^6} -\frac{(11c+656)}{6(c+24)} \frac{C_{44}^4 C_{46}^8}{C_{44}^6} \\
C_{66}^{[44]} & = \frac{784(c^2-172c+196)}{3(c+24)(2c-1)(7c+68)} \frac{\left(C_{44}^4\right) C_{46}^6}{\left(C_{44}^6\right)^2} -\frac{(11c+656)}{6(c+24)} \frac{\left(C_{44}^4\right) C_{46}^{[44]}}{C_{44}^6} \\
& -\frac{224(c^2-172c+196)}{(2c-1)(7c+68)^2} \frac{\left(C_{46}^6\right)^2}{\left(C_{44}^6\right)^2} +\frac{112(11c+656)}{9(c+24)^2} \frac{\left(C_{44}^4\right)^2}{\left(C_{44}^6\right)^2} +\frac{4(4c+61)}{(7c+68)} \frac{C_{46}^6 C_{46}^{[44]}}{C_{44}^6} \\
C_{66}^{[44]^{(2)}} & = \frac{1960(47c-614)(c^2-172c+196)}{3(c+24)(c+31)(2c-1)(7c+68)(55c-6)} \frac{C_{44}^4 C_{46}^6}{\left(C_{44}^6\right)^2} +\frac{3}{4} \frac{C_{46}^8 C_{46}^{[44]^{(2)}}}{C_{44}^6} \\
& +\frac{5(c+76)(5c+22)(11c+232)}{12(c+24)(c+31)(55c-6)} \frac{C_{44}^4 C_{46}^{[44]}}{C_{44}^6} \\
& -\frac{560(47c-614)(c^2-172c+196)}{(c+31)(2c-1)(7c+68)^2(55c-6)} \frac{\left(C_{46}^6\right)^2}{\left(C_{44}^6\right)^2} +\frac{280(11c+656)(47c-614)}{9(c+24)^2(c+31)(55c-6)} \frac{\left(C_{44}^4\right)^2}{\left(C_{44}^6\right)^2} \\
C_{66}^{10} & = \frac{3}{4} \frac{C_{46}^8 C_{48}^{10}}{C_{44}^6} \\
C_{66}^{[46]} & = -\frac{56}{(c+24)} \frac{C_{44}^4}{C_{44}^6} +\frac{48(81c+1274)}{(7c+68)(13c+516)} \frac{C_{46}^6}{C_{44}^6} +\frac{3(13c+248)}{2(13c+516)} C_{46}^{[44]} +\frac{3}{4} \frac{C_{46}^8 C_{48}^{[46]}}{C_{44}^6} \\
C_{46}^{[46]^{(1)}} & = -\frac{140}{3(c+24)} \frac{C_{44}^4}{C_{44}^6} - \frac{20(113c-1338)}{(7c+68)(13c+516)} \frac{C_{46}^6}{C_{44}^6} +\frac{5(13c+918)}{4(13c+516)} C_{46}^{[44]} - \frac{5}{4} \frac{C_{46}^8 C_{46}^{[44]^{(1)}}}{C_{44}^6}
\end{align*}

\section{Quadratic basis in even spin $\mathcal{W}_\infty$}
\label{apquad}
The field $V_6$ expressed in terms of $U_j$ fields is
\begin{align*}
V_6 & = U_6 - (U_1 U_5) + (U_2 U_4) - \frac{1}{2} (U_3 U_3) +\frac{1}{4} (5-2N) (U_1'' U_3) -\alpha_0 (N-3) (U_1' U_4) \\
& +\alpha_0 (N-3) (U_1 U_4') +\frac{1}{4} (N-2) (U_2'' U_2) +\alpha_0 (N-3) (U_2' U_3) -\alpha_0 (N-3) (U_2 U_3') \\
& -\frac{1}{4} (U_1 U_3'') + \frac{1}{96} (N-1) (72 \alpha_0^2+8 \alpha_0^2 N^2-50 \alpha_0^2 N-N) (U_1^{(4)} U_1) \\
& +\frac{1}{12} \alpha_0 (N-2) (30 \alpha_0^2+4 \alpha_0^2 N^2-22 \alpha_0^2 N-N) (U_1^{(3)} U_2) \\
& -\frac{1}{16} (30 \alpha_0^2+4 \alpha_0^2 N^2-22 \alpha_0^2 N-1) (12 \alpha_0^2+4 \alpha_0^2 N^2-14 \alpha_0^2 N-1) (U_1'' U_1'') \\
& +\frac{1}{4} \alpha_0 (N-2) (30 \alpha_0^2+4 \alpha_0^2 N^2-22 \alpha_0^2 N+2 N-7) (U_1'' U_2') \\
& -\frac{1}{4} \alpha_0 (N-2) (30 \alpha_0^2+4 \alpha_0^2 N^2-22 \alpha_0^2 N-1) (U_1' U_2'') \\
& +\frac{1}{2} (30 \alpha_0^2+4 \alpha_0^2 N^2-22 \alpha_0^2 N-1) (U_1' U_3') \\
& -\frac{1}{12} \alpha_0 (N-2) (30 \alpha_0^2+4 \alpha_0^2 N^2-22 \alpha_0^2 N-N) (U_1 U_2^{(3)}) \\
& +\frac{1}{4} (-30 \alpha_0^2-4 \alpha_0^2 N^2+22 \alpha_0^2 N+1) (U_2' U_2') \\
& +\frac{1}{24} \big[ -360 \alpha_0^4-12 \alpha_0^2-16 \alpha_0^4 N^4+144 \alpha_0^4 N^3+8 \alpha_0^2 N^3-476 \alpha_0^4 N^2 \\
& -42 \alpha_0^2 N^2+684 \alpha_0^4 N+60 \alpha_0^2 N-N \big] (U_1^{(3)} U_1') \\
& +\frac{1}{12} \alpha_0 (N-2) (30 \alpha_0^2+4 \alpha_0^2 N^2-22 \alpha_0^2 N-3) U_3^{(3)} \\
& +\frac{1}{48} (-30 \alpha_0^2+\alpha_0^2 N^3-7 \alpha_0^2 N^2+24 \alpha_0^2 N+1) U_2^{(4)} \\
& -\frac{1}{480} \alpha_0 (N-1) \big[1440 \alpha_0^4-168 \alpha_0^2+64 \alpha_0^4 N^4-576 \alpha_0^4 N^3-14 \alpha_0^2 N^3 \\
& +1904 \alpha_0^4 N^2 +52 \alpha_0^2 N^2-2736 \alpha_0^4 N+26 \alpha_0^2 N+5N \big] U_1^{(5)} \\
& -\alpha_0(N-3) U_5'
\end{align*}
OPE of field $V_4$ with itself is
\begin{align*}
V_4(z) V_4(w) & \sim -\frac{n}{2}\left(12\alpha_0^2+4\alpha_0^2n^2-14\alpha_0^2n-1\right) \left(2\alpha_0^2+4\alpha_0^2n^2-6\alpha_0^2n-1\right) \times \\
& \times \left(101\alpha_0^2+34\alpha_0^2n^2-117\alpha_0^2n-n-8\right) \frac{\mathbbm{1}}{(z-w)^8} \\
& + 4\alpha_0^2(n-1)(2n-5)(2n-1)\left(12\alpha_0^2+4\alpha_0^2n^2-14\alpha_0^2n-1\right) \frac{V_2}{(z-w)^6} \\
& + 2\alpha_0^2(n-1)(2n-5)(2n-1)\left(12\alpha_0^2+4\alpha_0^2n^2-14\alpha_0^2n-1\right) \frac{V_2^\prime}{(z-w)^5} \\
& -2(n-1)\left(12\alpha_0^2+4\alpha_0^2n^2-14\alpha_0^2n-1\right) \frac{(V_2 V_2)}{(z-w)^4} \\
& +(n-1)\left(-3\alpha_0^2+4\alpha_0^2n^2-8\alpha_0^2n-1\right) \left(12\alpha_0^2+4\alpha_0^2n^2-14\alpha_0^2n-1\right) \frac{V_2^{\prime\prime}}{(z-w)^4} \\
& +6\left(6\alpha_0^2+4\alpha_0^2n^2-12\alpha_0^2n+1\right) \frac{V_4}{(z-w)^4} \\
& -2(n-1)\left(12\alpha_0^2+4\alpha_0^2n^2-14\alpha_0^2n-1\right) \frac{(V_2^\prime V_2)}{(z-w)^3} \\
& +\frac{1}{3}(n-1)\left(-7\alpha_0^2+4\alpha_0^2n^2-6\alpha_0^2n-1\right) \left(12\alpha_0^2+4\alpha_0^2n^2-14\alpha_0^2n-1\right) \frac{V_2^{(3)}}{(z-w)^3} \\
& +3\left(6\alpha_0^2+4\alpha_0^2n^2-12\alpha_0^2n+1\right) \frac{V_4^\prime}{(z-w)^3} \\
& -(n-1)\left(12\alpha_0^2+4\alpha_0^2n^2-14\alpha_0^2n-1\right) \frac{(V_2^{\prime\prime} V_2)}{(z-w)^2} \\
& +\frac{1}{2}\left(-12\alpha_0^2-4\alpha_0^2n^2+14\alpha_0^2n+1\right) \frac{(V_2^\prime V_2^\prime)}{(z-w)^2} \\
& -\frac{4(V_2 V_4)}{(z-w)^2} - \frac{6V_6}{(z-w)^2} \\
& +\frac{1}{2} \left(-22\alpha_0^2+4\alpha_0^2n^2+1\right) \frac{V_4^{\prime\prime}}{(z-w)^2} \\
& +\frac{1}{24} \left(12\alpha_0^2+4\alpha_0^2n^2-14\alpha_0^2n-1\right) \times \\
& \times \left(22\alpha_0^2+8\alpha_0^2n^3-16\alpha_0^2n^2-14\alpha_0^2n-2n+1\right) \frac{V_2^{(4)}}{(z-w)^2} \\
& -\frac{1}{3}(n-1)\left(12\alpha_0^2+4\alpha_0^2n^2-14\alpha_0^2n-1\right) \frac{(V_2^{(3)} V_2)}{z-w} \\
& +\frac{1}{2}\left(-12\alpha_0^2-4\alpha_0^2n^2+14\alpha_0^2n+1\right) \frac{(V_2^{\prime\prime} V_2^\prime)}{z-w} \\
& +\frac{1}{60}\left(12\alpha_0^2+4\alpha_0^2n^2-14\alpha_0^2n-1\right) \left(15\alpha_0^2+4\alpha_0^2n^3-6\alpha_0^2n^2-13\alpha_0^2n-n\right) \frac{V_2^{(5)}}{z-w} \\
& -\frac{2(V_2^\prime V_4)}{z-w} -\frac{2(V_2 V_4^\prime)}{z-w} +\alpha_0^2(3n-7)\frac{V_4^{(3)}}{z-w} -\frac{3V_6^\prime}{z-w}.
\end{align*}

\bibliography{winfev}

\end{document}